\documentclass[%
 reprint,
superscriptaddress,
showpacs,preprintnumbers,
nofootinbib,
 amsmath,amssymb,aps,
prd,
floatfix,
]{revtex4-1}

\usepackage{lipsum} 

\usepackage{multirow}

\usepackage{bm} 
\usepackage{physics}
\usepackage{comment}
\pdfoutput=1
\usepackage{amsmath,amsfonts,amsthm}
\usepackage{esdiff} 
\usepackage{booktabs}  
\usepackage{url}  
\usepackage[hidelinks]{hyperref}  
\usepackage{relsize}

\usepackage[caption=false]{subfig}

\usepackage[normalem]{ulem}

\usepackage[usenames,dvipsnames,svgnames,table]{xcolor}

\usepackage[export]{adjustbox}
\usepackage{graphicx}

\usepackage[sc]{mathpazo} 
\usepackage[T1]{fontenc} 
\usepackage{microtype} 

\usepackage{booktabs} 
\usepackage{float} 
\usepackage{hyperref} 

\usepackage{lettrine} 
\usepackage{paralist} 

\DeclareCaptionFormat{myformat}{#1#2#3\hrulefill}
\begin{document}

\title{Estimating {amplitude} of matter density fluctuations in solar and supernova models using neutrino flavor evolution} 
\author{Caroline Laber-Smith}\thanks{Corresponding author}
\email{labersmith@wisc.edu}
\affiliation{Department of Physics, University of Wisconsin: Madison, Madison, WI 53706, USA}
\author{Hansen Torres}
\email{htorre04@nyit.edu}
\author{Lily Newkirk}
\email{lnewkirk@nyit.edu}
\author{Dhriti Rathod}
\email{drathod@nyit.edu}
\affiliation{Department of Physics, New York Institute of Technology, New York, NY 10023, USA}

\author{A. Baha Balantekin}
\email{baha@physics.wisc.edu}
\affiliation{Department of Physics, University of Wisconsin: Madison, Madison, WI 53706, USA}
\author{Amol V.\ Patwardhan}
\email{apatwardhan@reed.edu} 
\affiliation{Department of Physics, New York Institute of Technology, New York, NY 10023, USA}
\affiliation{Department of Physics, Reed College, Portland, OR 97202, USA}

\date{\today}


\begin{abstract}

Neutrinos can undergo substantial flavor evolution between their production in astrophysical sources---such as the Sun and core-collapse supernovae---and their subsequent detection in terrestrial detectors. This flavor evolution is strongly influenced by the environment that the neutrinos interact with, making them useful astrophysical messengers capable of potentially carrying information about the properties of the source wherein they are produced. In this work, we apply the framework of statistical data assimilation (SDA) in order to ascertain the extent to which a neutrino signal at detection may contain information about matter density fluctuations along their path from the source. Using simplified models of neutrino flavor evolution and propagation in the sun and in a core-collapse supernova (CCSN), we find that the SDA method {is able to extract information about the amplitude of density fluctuations in both the solar and CCSN scenarios, with the method showing relatively greater reliability in the solar neutrino case. Nonetheless, even in the CCSN neutrino model, the method proved effective at high fluctuation amplitudes.}

\end{abstract}


\maketitle

\section{Introduction} \label{sec:intro}
 
Neutrino flavor transformations have been, and continue to serve as, useful probes of astrophysical environments such as the Sun~\cite{Bahcall:2004qv} and core-collapse supernovae (CCSN)~\cite{Mirizzi:2015eza}. In the solar case, earth-based detectors { of various types---radiochemical (Homestake, GALLEX, SAGE), water Cherenkov (Kamiokande, Super-Kamiokande, SNO) and liquid scintillator (Borexino, KamLAND)---}have already amassed a large amount of neutrino data over the past few decades~\cite{Davis:1968cp, Davis:1978ew, Cleveland:1998nv, Kamiokande-II:1990wrs, SAGE:1994ctc, GALLEX:1994rym, SNO:2001kpb, SNO:2002hgz, SNO:2002tuh, SNO:2003bmh, SNO:2008gqy, SNO:2011hxd, Tolich:2012jh, Bellerive:2016byv, Borexino:2007kvk, Borexino:2008fkj, BOREXINO:2018ohr, BOREXINO:2021efb, BOREXINO:2022abl, BOREXINO:2023ygs, KamLAND:2011fld, KamLAND:2014gul, Jelley:2009, Abe:2024}. The solar environment is also more amenable to using this data as a sensitive probe, since the underlying physics{---of neutrino flavor transformation as well as that of the various environmental factors---}is much better understood. 

In the case of CCSN, however, existing data amounts to only a handful of neutrinos ($\sim 20$) from a single event (SN1987A)~\cite{Hirata:1987, Svoboda:1987, Alekseev:1987}, which leaves a lot of room for interesting physical phenomena in these environments which are yet essentially unconstrained. {Moreover, the environment itself is much more dynamically complex, both within the neutrino physics realm and otherwise. Due to the large neutrino densities near a CCSN core, neutrino-neutrino ($\nu$--$\nu$) interactions gain prominence, leading to a rich variety of collective phenomena in flavor space driven by coherent and collisional effects~\cite{duan2010collective, Mirizzi:2015eza, Chakraborty:2016yeg, Tamborra:2020cul, Capozzi:2022slf, Richers:2022zug, Volpe:2023met, Johns:2025mlm}. Even aside from neutrino flavor transformations, faithfully modeling the supernova environment has for years proven to be a highly complex theoretical and computational challenge, due to the underlying rich physics and variety of length and time scales in the problem~\cite{Janka:2017vcp, Mezzacappa:2020oyq, Burrows:2020qrp}.}
Nonetheless, the much more potent present-day and near future neutrino detectors offer a much greater promise that a future galactic (or near-galactic) CCSN event could yield orders of magnitude better data, allowing us to probe much deeper into the physics of these cataclysmic explosions~\cite{hyperK_2018,DUNE_Abi_2021,adam2015junoconceptualdesignreport}.
 
The plethora of interesting physical phenomena in these environments includes the possibility of irregularities in the matter profiles (density, temperature, composition, etc.) due to various effects such as convection, turbulence, magnetic fields, and rotation, which are sometimes neglected in simplified models of these environments. {Previous literature has demonstrated that such matter density fluctuations can significantly alter neutrino flavor evolution beyond the standard adiabatic MSW (Mikheyev-Smirnov-Wolfenstein~\cite{Wolfenstein:1977ue, mikheev1985resonance,Mikheev:1986}) picture~\cite{Sawyer:1990tw, Loreti:1994ry, Loreti:1995ae, Balantekin:1996pp, Schaefer:1987fr, Krastev:1989ix, Haxton:1990qb, Nunokawa:1996qu, Burgess:1996mz, Akhmedov:1999ty, Friedland:2006ta, Fogli:2006xy, Gava:2009pj, Choubey:2007ga, Kneller:2010ky, Kneller:2010sc, Kneller:2012id, Lund:2013uta, Patton:2013dba, Patton:2014lza, Yang:2015oya, Bhattacharyya:2025gds, Ma:2018key, Cherry:2011fm, Abbar:2020ror, Sigl:2021tmj, Mukhopadhyay:2023tsc, Reid:2011zz, Borriello:2013tha}. Early works demonstrated that even small-amplitude stochastic fluctuations can induce decoherence and suppress flavor conversion~\cite{Sawyer:1990tw, Loreti:1994ry, Loreti:1995ae, Balantekin:1996pp, Schaefer:1987fr, Krastev:1989ix, Haxton:1990qb}, while later studies extended this to supernova environments where shock waves and turbulence can produce phase effects, spectral splits, and non-adiabatic transitions~\cite{Nunokawa:1996qu, Burgess:1996mz, Akhmedov:1999ty, Friedland:2006ta, Fogli:2006xy, Gava:2009pj}. These studies show that the impact of fluctuations depends sensitively on their amplitude, correlation length, and location relative to resonance regions, with certain regimes enhancing flavor conversion and others washing out oscillation signatures~\cite{Choubey:2007ga, Kneller:2010ky, Kneller:2010sc, Kneller:2012id, Lund:2013uta, Patton:2013dba, Patton:2014lza, Yang:2015oya}. In core-collapse supernovae, the interplay between matter fluctuations and collective neutrino oscillations further complicates the dynamics, leading to nonlinear and potentially chaotic behavior~\cite{Bhattacharyya:2025gds, Ma:2018key, Cherry:2011fm, Abbar:2020ror, Sigl:2021tmj, Mukhopadhyay:2023tsc}. Additional work has highlighted how these effects manifest in observable signals and the importance of accurately modeling them for interpreting data~\cite{Reid:2011zz, Borriello:2013tha}.
}

In this paper, we take a statistical data assimilation (SDA) approach~\cite{kalnay2003atmospheric, evensen2009data, betts2010practical, abarbanel2013predicting, abarbanel2017unifying, rey2014accurate} to examine the degree to which limits on these irregularities could be placed purely from neutrino flavor observations.
This approach, developed originally for weather forecasting~\cite{kalnay2003atmospheric}, is well-suited to problems where we want to use sparse data to inform a dynamic model. It has been applied across fields as diverse as geophysical modeling~\cite{whartenby2013number,an2017estimating}, neurobiology~\cite{schiff2009kalman, abarbanel2011dynamical, toth2011dynamical,kostuk2012dynamical,hamilton2013real,meliza2014estimating,nogaret2016automatic, abarbanel2017unifying, breen2016hvc, armstrong2020statistical}, spacecraft orbital dynamics~\cite{betts2010practical}, exoplanet modeling~\cite{Madhusudhan2018}, and solar cycle prediction~\cite{Kitiashvili2008,kitiashvili2020}.  This potentially makes it a good fit for neutrino oscillations, particularly with CCSN neutrinos, {wherein the different neutrino energy/angular modes are coupled to each other by the nonlinear terms in the Hamiltonian, facilitating flow of information across the system in the SDA framework}. 

{ SDA-based methods have been tested previously with simplified models of solar~\cite{laber2023inference, laber2024v} as well as CCSN neutrino flavor evolution~\cite{armstrong2017optimization, Armstrong:2020gxk, Armstrong:2022tsj, Newkirk:2024mnu}. For solar neutrinos, }
 this problem initially used a simpler, more well-studied test case.  The objective was to use SDA to infer the presence of neutrino-electron interactions and to compare the performance of SDA to typical forward integration techniques~\cite{laber2023inference}.  For this, measurements of solar neutrino flux from Borexino and SNO were incorporated, and incomplete information was provided about the model dynamics. Subsequently, this method was extended by allowing the SDA procedure to experiment with different matter density profiles, testing the adiabaticity of flavor evolution and the expected electron density at the center of the Sun~\cite{laber2024v}. {Likewise, using simplified models for interacting neutrinos in a CCSN environment, the SDA procedure was put to the test over its sensitivity to matter density profiles---both using analytic matter profiles~\cite{armstrong2017optimization, Armstrong:2020gxk} and numerical outputs from simulations~\cite{Newkirk:2024mnu}---as well as to infer the presence of bipolar oscillations in the CCSN envelope~\cite{Armstrong:2022tsj}. 
 
 However, none of these previous studies have yet probed the extent to which this procedure can be amenable to identifying random irregularities in such environments using real or anticipated neutrino signals in terrestrial detectors. The current study offers a first exploratory step in that particular direction.} {The manuscript is organized as follows: in Sec.~\ref{sec:model}, we describe the underlying physical models of our simplified systems in the solar and the CCSN cases, along with details of how the irregularities in the matter profile are incorporated (Sec.~\ref{sec:noise}). Section~\ref{sec:methods} lays out the details of the SDA framework. Results, and discussion thereof, are presented in Sec.~\ref{sec:results}, and we conclude in Sec.~\ref{sec:conclusion}.

}

\section{Model} \label{sec:model}

We modeled neutrino flavor evolution using a two-flavor approximation, where neutrinos are considered to be {a linear combination of the} electron flavor and $x$ flavor {(particular superposition of $\mu$ and $\tau$)}. This can be {implemented in the solar case} using the mixing angle $\theta \simeq \theta_{12}$ and the mass-squared difference $\delta m^2 \simeq \delta m^2_{12}$.  The two-flavor approximation is an effective one in the sun due to the small value of $\theta _{13}$ and the well-separated hierarchy of the two mass-squared differences~\cite{Lim:2002iz, Balantekin:2003dc}. { In the CCSN case, nonlinear effects likely make the two-flavor approximation less robust~\cite{Dasgupta:2007ws, Esteban-Pretel:2007ncu, Duan:2008za, Mirizzi:2011zza, Dasgupta:2010ae, Dasgupta:2010cd, Doring:2019axc, Chakraborty:2019wxe, Shalgar:2021wlj}---however, in the interest of simplicity, and to facilitate comparison with the solar results, we persist with the two-flavor toy model even in the CCSN case.} 

Other specifics of this approach varied between {solar and CCSN neutrinos}. In the solar environment, the dominant neutrino refractive potential arises from coherent forward scattering of neutrinos due to interactions between neutrinos and ordinary matter inside the Sun~\cite{Wolfenstein:1977ue, mikheev1985resonance, Mikheev:1986}. {In the Sun, unlike in a CCSN, neutrino fluxes are not high enough for the $\nu$--$\nu$ interactions to contribute meaningfully to the refractive potential. As a result, each energy mode in the emergent neutrino spectrum evolves independently of the others. This provides us with the discretion to choose the number of energy modes in our implementation, without it affecting the underlying model dynamics. {For our solar neutrino computations in this work, we adopted four energy modes, with energies chosen to mirror those from a recent analysis~\cite{Denton:2025cbo} that used extracted survival probability values from SNO and Super-Kamiokande data.}
  
{ In the CCSN environment, neutrino fluxes are sufficiently high for the $\nu$--$\nu$ interactions to be a significant component of the refractive potential. As a result, each neutrino mode no longer evolves in flavor independently of the others, making it a coupled, nonlinear problem wherein the model dynamics can vary based on the number of energy modes in the implementation.} {Nonetheless, here we sought} to provide a proof-of-concept illustration of the sensitivity of the SDA procedure (details in Sec.~\ref{sec:methods}) to extract information about matter density fluctuations in a CCSN environment using simulated neutrino data. To this end, we implement a minimal model of an interacting two-flavor neutrino system with two energy modes. These modes interact with each other and with a matter background. The angular dependence {as well as the explicit time dependence} of the flavor evolution are neglected. In the CCSN case, we parameterize the two-flavor mixing using the mass-squared difference $\delta m^2 \simeq \delta m^2_{13}$ and a mixing angle $\theta \simeq \theta_{13}$. The primary reason to switch to two modes in the CCSN case, compared with 
four in the solar case was logistical---the CCSN model requires a much finer grid spacing, coming at a large computational cost to the SDA procedure, which then had to be partly offset by reducing the number of modes. The parameter values the solar and CCSN computations are given in Tables~\ref{table:Energies} and \ref{table:Known}, respectively. %

\setlength{\tabcolsep}{5pt}
\begin{table}[htb]
\small
\centering
\begin{tabular}{|l | r | } \toprule
\hline
\textit{Parameter} (unit) & \textit{Value}  \\\midrule \hline
Energies (MeV) &  7, 9, 11, 13 \\
$\delta m^2$ (eV$^2$) & $7.53\times10^{-5}$ \\
$\theta$ & 0.584 \\
\bottomrule \hline
\end{tabular}
\caption{Parameter values~\cite{Abe:2024,Super-Kamiokande:2023ahc} used for the solar neutrino model calculations. } 
\label{table:Energies}
\end{table}

\setlength{\tabcolsep}{5pt}
\begin{table}[htb]
\small
\centering
\begin{tabular}{|l |r |} \toprule
\hline
\textit{Parameter} (unit) & \textit{Value} \\\midrule \hline
 Energies (MeV) & 14, 11 \\ 
 $\delta m^2$ (eV$^2$) & $2.44\times10^{-3}$ \\
 $\theta$ & 0.1 \\
 $L_{\nu}$ (erg/s)& $1.23 \times 10^{52}$ \\
 $\langle E_{\nu} \rangle$ (MeV) & 13.5 \\
  $R_{\nu}$ (km) & $20$ \\
\bottomrule \hline
\end{tabular}
\caption{Parameter values~\cite{Abe:2024,Super-Kamiokande:2023ahc, Radice:2017ykv, Wang:2024tbv, Burrows2023} used for the CCSN neutrino model calculations.
}
\label{table:Known}
\end{table}

{In our models}, we represent neutrino flavor in each energy mode using polarization vectors $\vec{P}(E)$, defined in terms of the neutrino { flavor-space} density matrices $\rho(E)$ as follows:
{
\begin{equation}
    \rho(E) =  
    \begin{bmatrix}
    \rho_{ee}(E) & \rho_{ex}(E) \\
    \rho_{xe}(E) & \rho_{xx}(E)
    \end{bmatrix}
    = \frac12 \left[\mathbb{I} + \vec{\sigma} \cdot \vec{P}(E)\right].
\end{equation}
}
{Henceforth, for brevity, we shall drop the explicit energy labels, and instead denote each mode with energy $E_i$ using the corresponding polarization vector label $\vec{P}_i$. 
The density matrices are normalized to $\rho_{ee} + \rho_{xx} = 1$, and therefore,} the $z$-component of $\vec{P}(E)$ can be directly connected to the { probability $P_{\nu_e}$ of finding a neutrino in the electron flavor}:
\begin{equation}
    P_z = \rho_{ee} - \rho_{xx} = 2 \rho_{ee} -1 \equiv 2 P_{\nu_e} - 1.
\end{equation}

For flavor evolution inside of the Sun, our model dynamics are described by:
{\begin{equation} \label{eq:modelsun}
    \dv{\vec{P}_i}{r} = \left(\Delta_i\vec{B}+V(r)\,\hat{z}\right)\cross \vec{P}_i.
\end{equation}  
}

The first term {in the parentheses is responsible for neutrino flavor oscillations in vacuum}, due to each neutrino flavor state being composed of a superposition of mass eigenstates. $\Delta_i$ and $\vec{B}$ can be expanded as:
\begin{equation}
\Delta_i = \frac{\delta m^2}{2E_i}, \quad \vec{B} = \sin 2\theta \,\hat{x} - \cos 2\theta \,\hat{z}.
\end{equation}

Here, $\delta m^2$ and $\theta$ represent the mass-squared difference between neutrino mass eigenstates in vacuum, and the mixing angle in vacuum, respectively. See Table \ref{table:Energies} for the values used. The second term in the parentheses represents matter effects due to interactions between neutrinos and the background electrons.  Here, $V(r)$, {representing neutrino forward scattering off of the background medium,} can be defined in terms of the electron density $n_e(r)$ and the Fermi coupling constant $G_F$:
\begin{equation}
    V(r) = \sqrt{2} G_F n_e(r).
\end{equation}

For the {baseline} electron density $n_e$ {(without fluctuations)}, we used a table of values from the {BS05(OP)} standard solar model \cite{Bahcall_2005, Bahcallweb}. 
{{Since our flavor evolution computations required a finer radial grid (i.e., higher step count)} than was available, we linearly interpolated between the provided values of $n_e(r)$ as needed.}

Describing the neutrino flavor dynamics inside of a CCSN, where neutrino number densities are much higher, requires an additional term representing {$\nu$--$\nu$ forward and exchange scattering~\cite{Pantaleone:1992eq, Pantaleone:1992xh}}.  For each of our energy modes described by $\vec{P}_i$, the evolution equations become:
{\begin{equation} \label{eq:modelccsn}
  \diff{\vec{P}_{i}}{r} = \left(\Delta_i \vec{B} + V(r) \hat{z} 
  +\mu(r) \sum_{j\neq i} \vec{P}_j \right) \times \vec{P}_i.
\end{equation}
} 

{The potential function $V(r)$ here (without fluctuations) came from a one-dimensional simulation of a CCSN, with a 9.6\,$M_\odot$ progenitor with metallicity of $Z = 10^{-4}$~\cite{Radice:2017ykv, Wang:2024tbv, Burrows2023}. In particular, we used a snapshot of the simulation at 0.01\,s post core-bounce. As with the solar case, we used linear interpolation in between the available density values from the simulation snapshot.}

The $\nu$--$\nu$ interaction is encoded in a potential function: 
\begin{equation} \label{eq:Vnu}
  \mu(r) = \frac{L_{\nu}}{4\pi \langle E_{\nu} \rangle R_{\nu}^2} \sqrt{2} G_F \left(1 - \sqrt{ 1 - \left(\frac{R_\nu}{r}\right)^2}\right)^2.
\end{equation}
{which emerges from the \lq neutrino bulb\rq\ model in the single-angle approximation~\cite{duan2006simulation,duan2010collective}.} Values used for the parameters in this model are shown in Table \ref{table:Known}.

\subsection{Implementation of density fluctuations}  \label{sec:noise}

{To appraise the impact of matter density fluctuations on the final flavor evolution outcomes, we perturbed the matter potential $V(r)$ in both the solar and the CCSN cases with an added noise: 
\begin{equation} \label{eq:Vrwithnoise}
    V(r) \rightarrow V(r) \left[1+{\Sigma(r)}\right].
\end{equation}
}

{The noise function $\Sigma(r)$ was generated by assigning a random number sampled from a Gaussian distribution---with mean 0, and standard deviation $\sigma$---at each point of our discretized grid, then filtering out fluctuations above an adjustable minimum frequency. We set the minimum fluctuation length to 100 steps by applying a low-pass filter to remove all fluctuations on smaller length scales. Since 100 steps corresponds to 2\% of the Nyquist frequency (0.5 cycles/step), and because Gaussian white noise has uniform power spectral density, the variance of the resulting filtered noise is 2\% of the variance of the original Gaussian.
The resulting standard deviation, $\tilde\sigma$, of the filtered noise is therefore $14.14\%$ ($= 1/\sqrt{50}$) of the Gaussian standard deviation $\sigma$. We henceforth use $\tilde\sigma$ as a measure of the relative strength of the density fluctuations, and refer to it as the \textit{noise amplitude}.

The rationale behind choosing 100 steps as the minimum fluctuation length was to allow the noise {to be fully resolved by our step size, but to have it be faster than the slow, MSW-induced evolution.}}
As a reference point for the {length scale} of these density fluctuations, we can compare them to other relevant length scales in our model dynamics.  Our grid size $\delta r$, flavor oscillation length $L_{\text{osc}}$, minimum fluctuation length $L_{\text{Noise}}$, and the scale height $H$ of the matter potential (without noise) follow this hierarchy:
\begin{equation}
    \delta r \ll L_{\text{osc}} \sim L_{\text{Noise}} \ll H. 
\end{equation}

{Here, $L_\text{osc} = 4\pi E_\nu/\delta m^2_\text{eff}$, for neutrino energy $E_\nu$, and where $\delta m^2_\text{eff}$ is the in-medium mass-squared difference between the two instantaneous eigenstates of the neutrino Hamiltonian. The scale height of the matter potential can be calculated as $H = [d \log n_e/dr]^{-1}$.}

{For both solar and CCSN neutrinos}, we {tested the efficacy of our procedure across a range of noise amplitudes}. {We present our main results with standard deviations $\sigma$} of the sampled Gaussian noise chosen to have values of $2\%, 10\%, \text{and }30\%$ for generating the simulated measurements that we provide to the SDA procedure (see Sec. \ref{sec:methods}).  The corresponding {amplitudes} of the filtered noise are shown in Table \ref{table:SearchRanges}, {second column}.

\setlength{\tabcolsep}{5pt}
\begin{table}[htb]
\small
\centering
\begin{tabular}{|l |l |l |} \toprule
\hline
\textit{Sampled Gaussian $\sigma$} & \textit{{ Filtered} Noise $\tilde\sigma$} & \textit{$\tilde\sigma$ Search Range}  \\\midrule \hline
 $2\%$ & $0.283\%$ & $0 \text{--} 2.828\%$ \\
 $10\%$ & $1.414\%$ & $0 \text{--} 14.142\%$ \\
 $30\%$ & $4.243\%$ &  $0 \text{--} 14.142\%$ \\
\bottomrule \hline
\end{tabular}
\caption{Standard deviations of the unfiltered and filtered noise used for generating the simulated measurements, along with the permitted search ranges for the {filtered noise amplitude} given to the SDA procedure in each case.} 
\label{table:SearchRanges}
\end{table}

\section{Methods} \label{sec:methods}

\subsection{Inference process}

Statistical data assimilation (SDA) is an inference technique which incorporates information about model dynamics in combination with measurements.

For our implementation, we treat a system (in this case, neutrino oscillations) as a set of $D$ state variables $x_a(r)$, with behavior described by corresponding differential equations. We also incorporate parameters $\bm{p}$, which are constant over the region of interest but effect model dynamics.  These parameters can be adjusted within a defined range to find values which allow for the best match to measurements and model dynamics.

For computational purposes, we describe the model dynamics via a discrete model $F_a$ such that:
\begin{equation}
\dv{x_a(r_{n})}{r} = F_a(\bm{x}(r_n),\bm{p}).
\end{equation}

In our case, these would correspond to Eqs.~\eqref{eq:modelsun} (solar) or \eqref{eq:modelccsn} (CCSN). The model is imposed via a cost function $A_0$, also referred to as \textit{action}.  $A_0$ can be broken into two separate components:
\begin{equation} \label{eq:action}
A_0 = R_f A_{\text{model}} + R_m A_{\text{meas}}.
\end{equation}

$A_{\text{model}}$ penalizes deviation from the model dynamics $F_a$, while $A_{\text{meas}}$ penalizes difference from any given measurements.  These are implemented as: 
{\begin{widetext}
\begin{equation} \label{eq:actionlong}
\begin{split}
A_\text{model}=&\frac{1}{{N}D}	\mathlarger{\sum}_{n \in \{\text{odd}\}}^{N-2} \, \mathlarger{\sum}_{a=1}^D \\ 
   & \Bigg[ \left\{x_a(r_{n+2}) - x_a(r_n) - \frac{\delta r}{6} [F_a(\bm{x}(r_n), \bm{p}) + 4F_a(\bm{x}(r_{n+1}),\bm{p}) + F_a(\bm{x}(r_{n+2}),\bm{p})]\right\}^2 \\
   & + \left\{ x_a(r_{n+1}) - \frac12 \left(x_a(r_n)+x_a(r_{n+2})\right) - \frac{\delta r}{8} [F_a(\bm{x}(r_n),\bm{p}) - F_a(\bm{x}(r_{n+2}),\bm{p})]\right\}^2 \Bigg] \\
  A_{\text{meas}} =& \frac{1}{N_{\text{meas}}} \mathlarger{\sum}_{r_m \in \{\text{meas}\}} \, \mathlarger{\sum}_{l=1}^L  \, \big[y_{l}(r_m) - h_{l,m}(\bm{x}(r_m)) \big]^2 
\end{split}
\end{equation}
\end{widetext}}

In this definition of $A_{\text{model}}$, $N$ and $D$ refer to the number of discrete radial steps used and the number of state variables respectively.  Here, $\delta r = r_{n+2} - r_n$, i.e., twice the size of each radial step.  The definition of $A_\text{model}$ comes out of treating state variable evolution as a Markov chain, where the probability associated with each step can be determined using the model dynamics $\bm{F}$.  For detailed derivation of this term, refer to Ref.~\cite{armstrong2017optimization}.

For the measurement component, $\vec{y}$ refers to the set of measurements imposed.  $N_{\text{meas}}$ counts the number of these measurements.  In this case, we used generated measurements---refer to Sec.~\ref{sec:constr} for further details.  The function $h(\bm{x})$ is used to transform state variables into measurable quantities.  

To find the best fit to the underlying dynamics and constraints, we search for paths {$\{ \bm{x}(r_0), \ldots, \bm{x}(r_N), \bm{p} \}$ in state space} that minimize $A_0$.  This is done via standard numerical optimization techniques.  However, as $A_0$ is dependent on the value of $\bm{x}$ at every location{, and on the parameters $\bm{p}$,} the task of finding the global minimum of $A_0$ starting from completely random paths is too computationally complex to be feasible.  To make this more approachable we use a simulated annealing procedure, where the problem is repeatedly solved at increasing levels of complexity. At each new step, the previous solutions act as our initial guesses for the next optimization task.  

To implement this simulated annealing, the scaling factors $R_m$ and $R_f$ are used to control the relative weights of the model and measurement terms.  We begin with $R_f$ set to a very small value $R_{f,0}$ such that the vast majority of the cost function comes from measurement error.  We find multiple paths that approximately minimize $A_0$ with these factors and then we increase the weight of the model term following:
 \begin{equation} \label{eq:Rf}
    R_f = R_{f,0} \alpha^\beta.
\end{equation}

Here, $\alpha$ is a chosen constant value that can be adjusted to fit the problem.  $\beta$ starts at $0$ and is incremented by 1 with each iteration. To carry out this process we use the Python interface \texttt{minAone}~\cite{minAone}, which generates C++ code to perform optimization via the Interior Point Optimizer library (\texttt{IPOPT})~\cite{ipopt_article,ipopt_repo,wachter2009short}.  

\setlength{\tabcolsep}{5pt}
\begin{table}[htb]
\small
\centering
\begin{tabular}{|c|c|c|c|}
     \hline Model & Noise Amplitude & $\log_{10} R_{f,0}$& $\alpha$  \\ \hline 
     \multirow{3}{*}{Solar} &  $0.283\%$ & 2 & 1.5\\
      & $1.414\%$ & 2 & 1.5 \\
      & $4.243\%$ & 2 & 1.5 \\ \hline
     \multirow{3}{*}{CCSN}  &  $0.283\%$ & 
     $-1$ & 2\\
     & $1.414\%$ & $-1$ & 2 \\
     & $4.243\%$ & 1 & 2 \\\hline
\end{tabular}
\caption{Values of annealing parameters $R_{f,0}$ and $\alpha$, defined in Eq.~\eqref{eq:Rf}, used for the different SDA runs, organized by model (Solar/CCSN) in column 1  and the true noise amplitude in column 2. The different choices of annealing parameters in each case were necessary to reduce the likelihood of the paths falling into and becoming trapped in local minima in state space over the course of annealing.} 
\label{table:SearchRanges}
\end{table}

\subsection{Simulated measurements} \label{sec:constr}

To {generate simulated data for the inference procedure}, we used results from {forward-integrating the neutrino evolution equations in each case}. For the forward integration, the noise terms were explicitly added to $V(r)$ in accordance with Eq.~\eqref{eq:Vrwithnoise}, and with the amplitudes given in Table.~\ref{table:SearchRanges}.  {Each forward run} began with an initial state of $\vec{P} = \hat{z}$, representing pure electron flavor. { In the solar model, this represents regular emission which is $\nu_e$ dominated, whereas in a CCSN, such an initial condition would be characteristic of the deleptonization burst phase{---consistent with our usage of an early (0.01\,s) simulation snapshot of $V(r)$.}}

{For solar neutrinos}, we {defined a grid with} 121,901 steps, covering the distance between the center of the Sun and {a final radius of $ R_\odot/2 $, i.e., half the solar radius}.  
{The CCSN grid} covered a range of $r \in \left[1761.58,19939.83\right] \text{ km}${, which spans the MSW resonance region for the atmospheric mass-squared difference}, and we increased the step count to $10^6 + 1$.  This higher step count was needed to fully resolve the { much smaller oscillation length scales in this environment}, particularly near the inner section of the investigated region.

At the end of the region, we provided measurements corresponding to the average fraction of electron neutrinos that would be observed on Earth.  This measurement accounts for the observed neutrinos leaving the region of interest and traveling through a vacuum, where $n_e(r) \approx 0$ and neutrino density decreases such that $\nu$--$\nu$ interactions can be ignored. To do this, we take the results from forward integration and calculate {the electron flavor survival probability} $P_{\nu_e, \oplus}$ at the earth using a transformation function $h\big(\vec{P}\big)${, which also factors in the kinematic decoherence of the neutrinos along their propagation from the source to the earth}~\cite{laber2023inference, laber2024v, Dighe:1999id}:
\begin{equation}
    \label{eq:Psurv}
    2P_{\nu_e, \oplus} - 1 \equiv h\big(\vec{P}\big) \\
    =\cos^2(2\theta) P_{z,f} - \cos(2\theta) \sin(2\theta) P_{x,f}.
\end{equation}

{Here, $P_{z,f}$ and $P_{x,f}$ represent values of the polarization vector components in the sufficiently dilute regime where the matter and/or neutrino potentials have become negligible. The values of $P_{\nu_e,\odot}$ thus calculated from the forward integration runs are provided as simulated measurements to the SDA procedure, serving as constraints on linear combinations of $P_z$ and $P_x$ state variables near the outer endpoint of our domain. Additionally, we also impose $P_z = 1$ as a measurement at the inner boundary, in both the solar in CCSN models.}\footnote{In imposing $P_z = 1$ at the inner boundary of the CCSN grid, there is the implicit assumption that no flavor transformations occur between the neutrinosphere and this boundary. This isn't unreasonable during the the deleptonization burst stage, where the overwhelming preponderance of $\nu_e$ over all other species results in a suppression of collective oscillations.}
We penalize deviation between our measurements and $h\big(\vec{P}\big)$ at the last 1000 steps,  weighting the penalty at each step to match the overall weight of the {inner endpoint} constraint.  This penalty is imposed with the measurement cost function $A_{\text{meas}}$, as described in Eq.~(\ref{eq:actionlong}).

\subsection{Noise {amplitude} estimation}

Our goal was to see what information we could infer about the {amplitude} of density fluctuations using only these constraints. 

We tried two setups: first, we gave the SDA procedure the shape of the noise profile used for forward integration, but allowed it to {infer} the {amplitude $\tilde\sigma$} of the noise.  The model attempts to select a noise {amplitude} which allows it to find a state variable solution with lowest action, {seeking maximum simultaneous compatibility with both the model dynamics and the constraints}.

For the second setup, we provided the SDA procedure with {different} noise profiles {than the ones used by the forward integration code to generate the simulated constraints}. 
The {noise amplitude $\tilde\sigma$} in the SDA procedure was again kept as a parameter to be {inferred}.

In both cases we used an allowed search range for $\tilde\sigma$ going from no noise up to 10 times the value used for forward integration, capped at 14.142\% {(corresponding to $100\%$ standard deviation in the unfiltered noise)} for the 4.243\% noise amplitude case.  Our exact search ranges {for each case} are shown in Table \ref{table:SearchRanges}, {right column}. {For each scenario, the SDA procedure used 10 paths, randomly initialized in state space at the onset of the first annealing step.}

\section{Results} \label{sec:results}

{Here we present the results of the SDA procedure for the solar (Sec.~\ref{sec:solar}) and CCSN (Sec.~\ref{sec:ccsn}) neutrino models. In each case, we first show results from the trials where the SDA was provided with the same noise profile shapes as the ones used for simulated data generation, followed by the results from trials where the SDA used different randomized noise profile shapes.
In Sec.~\ref{sec:statevar}, the state variable evolution comparisons between forward integration (true) and SDA (inferred) are shown for both the solar and CCSN results.}

\subsection{Solar Neutrinos} \label{sec:solar}

{First, we examine the solar neutrino case. For the results shown in Fig. \ref{fig:solar4PsurvIPOPT}, the SDA was provided with the same noise profile shapes as the forward integration. The right panels show the noise {amplitude} estimates obtained by 10 randomly initialized paths over the course of annealing, for each of the the true noise amplitudes used for generating the simulated data. The evolution of the action levels associated with each path [Eqs.~\eqref{eq:action}--\eqref{eq:actionlong}] across the annealing steps is shown in the corresponding left panels. The dashed lines in the right panels, along with the labels on top of each row of sub-figures, indicate the true value of the noise amplitude in each case. The action levels after the annealing converges can be used as a diagnostic for identifying the paths that find the true solution (or ones degenerate therewith), which becomes important when the true solution is not known \textit{a priori}~\cite{Armstrong:2020gxk}.}

\begin{figure}[htb]
    \includegraphics[width=\linewidth]{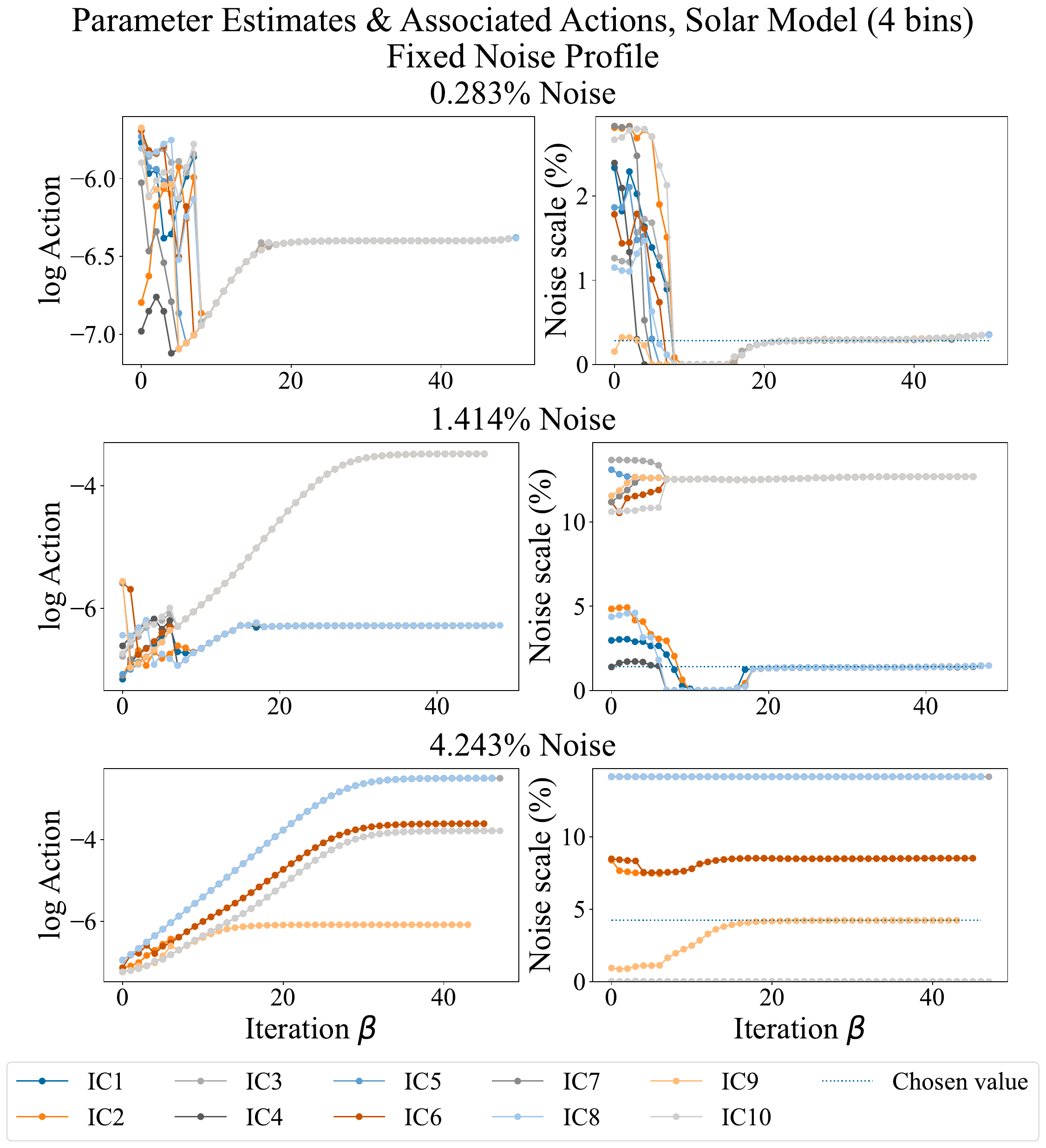}
    \caption{{Noise amplitude} parameter estimates {(right)} and corresponding action curves {(left)} {versus annealing parameter $\beta$}, for the solar neutrino model, across 10 sets of randomized initial conditions (ICs) for the paths.  The dotted lines on the right show the noise {amplitudes} in each case used to generate the simulated measurements for the SDA procedure. {The panels are arranged from top to bottom in increasing order of this true noise amplitude.} Measurements were provided of $P_z$ for each energy mode at $r=0$, and of $P_{\nu_e,\oplus}$ (expressed as linear combinations of $P_z$ and $P_x$) at the final 1000 steps.  {For the trials shown here, the same noise profile shape was shared between the forward simulation and SDA.}
    }
    \label{fig:solar4PsurvIPOPT}
\end{figure}

\begin{figure}[htb]
    \includegraphics[width=\linewidth]{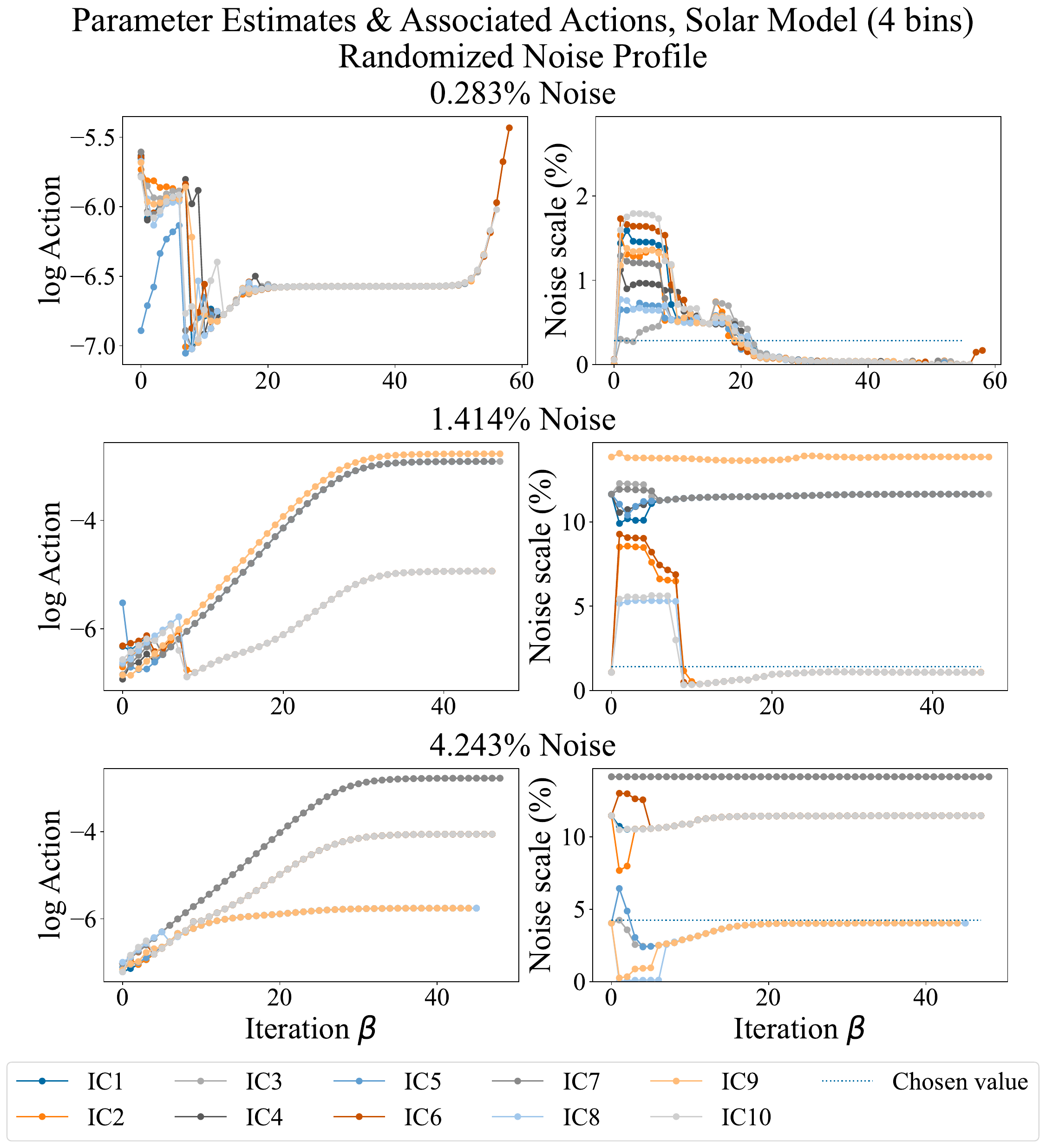}
    \caption{Results from the same procedure as Fig.~\ref{fig:solar4PsurvIPOPT} for the solar neutrino model, but with different randomized noise profiles provided for SDA than the ones used by forward integration to generate simulated measurements.}
    \label{fig:solar4PsurvDiff}
\end{figure}

In the lowest noise case (0.283\%) for solar neutrinos, {the parameter estimates from \textit{all} 10 paths converged to the true noise amplitude after about 20 annealing iterations. This ended up being the case even though the paths initially exhibited a wide range of parameter estimates during the early annealing stages (small $\beta$ values). Correspondingly, all these paths also attained nearly identical asymptotic action levels after $\beta \geq 20$.} 

The results for the $\tilde\sigma = 1.414\%$ case showed a similar trend for four of the paths, with each of them converging to a noise amplitude estimate near that used for forward integration, and displaying a plateau in action with the lowest level among the paths considered.  The remaining six paths converged to a much higher value for noise {amplitude}, and had action values over two orders of magnitude higher{, indicating that those paths found a worse match between model and data (due to their incorrect noise amplitude guesses)---a telltale sign of getting stuck in a local minimum in state space.}

For the $\tilde\sigma = 4.243\%$ case, the paths again showed a spread in final inferred noise {amplitude}s.  Only one path predicted an amplitude close to the true value, but this path had the lowest action by a margin of two orders of magnitude.  Paths which chose noise {amplitude}s below or above this all had significantly higher deviation from the provided constraints. {In both the $\tilde\sigma = 1.414\%$ and $ 4.243\%$ cases, it is thus evident that the action levels could be reliably used as a diagnostic for identifying whether the true solution had been found (by some subset of paths).}

For the case where the SDA procedure is given a different randomized noise profile than the one used to generate measurements via forward integration, the results are shown in Fig. \ref{fig:solar4PsurvDiff}.   {For the lowest noise {amplitude} case, none of the ten paths were able to converge to the true $\tilde\sigma$ value of 0.283\%, but after $\beta \geq 20$, they all were driven towards preferring a nearly zero noise amplitude.
For the higher noise amplitudes, similar patterns were observed as in Fig.~\ref{fig:solar4PsurvIPOPT}. In each case, a fraction of the paths converged to the true $\tilde\sigma$ value, and those paths also corresponded to the lowest action levels on the plateau.}

\subsection{CCSN Neutrinos} \label{sec:ccsn}

{The results from the CCSN neutrino model SDA computations are displayed in Fig.~\ref{fig:ccsnPsurvIPOPT} for the trials with the same noise profile shared between SDA and forward integration, and in Fig.~\ref{fig:ccsnPsurvDiff} for the trials which used different noise profile shapes for SDA.} 

\begin{figure}[htb]
    \includegraphics[width=\linewidth]{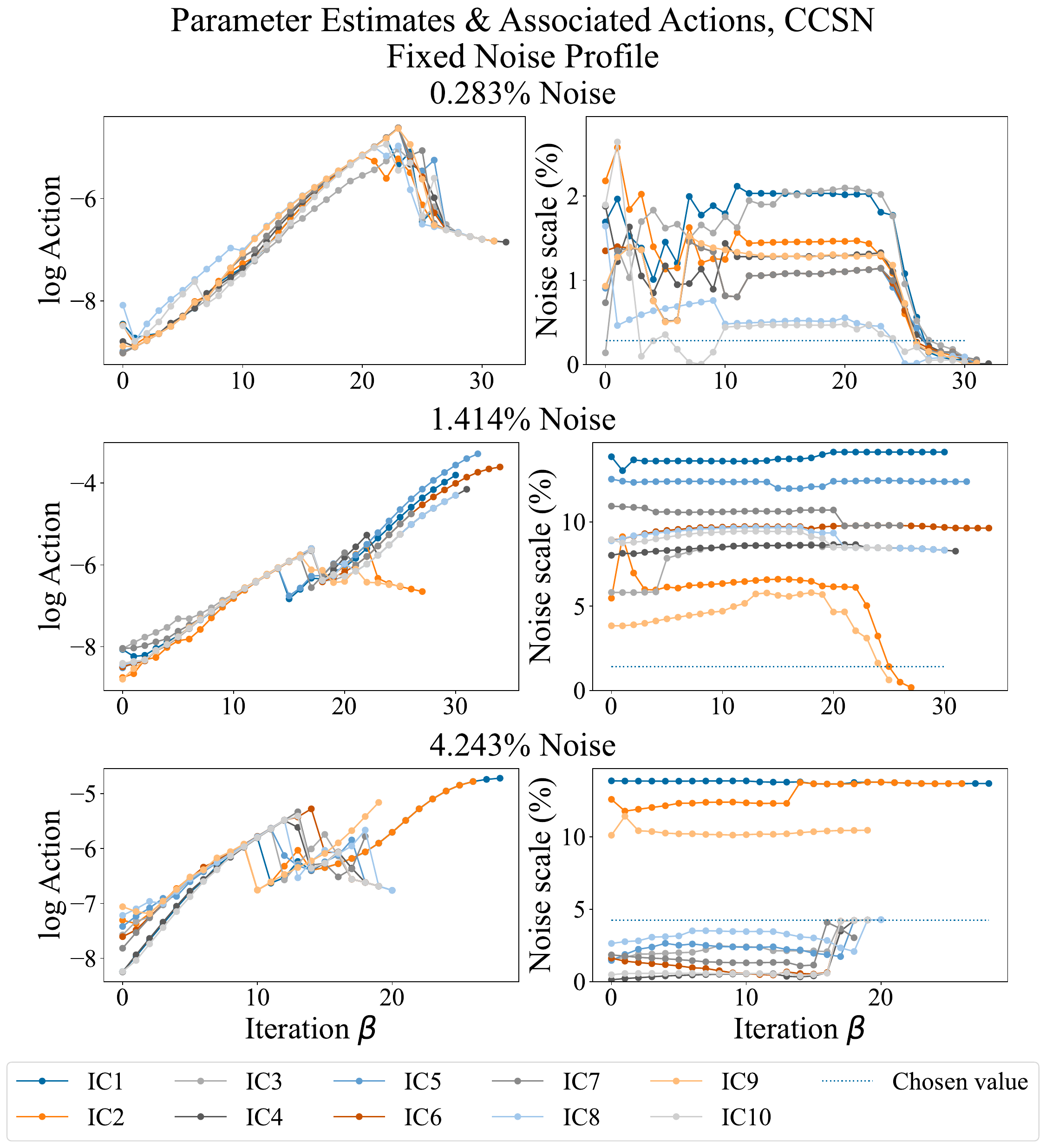}
    \caption{Results of the SDA procedure for the CCSN neutrino model. The quantities shown are organized in the same way as Fig.~\ref{fig:solar4PsurvIPOPT}. For the trials shown here, the same noise profile shape was shared between the forward simulation and SDA.} 
    \label{fig:ccsnPsurvIPOPT}
\end{figure} 

\begin{figure}[htb]
    \includegraphics[width=\linewidth]{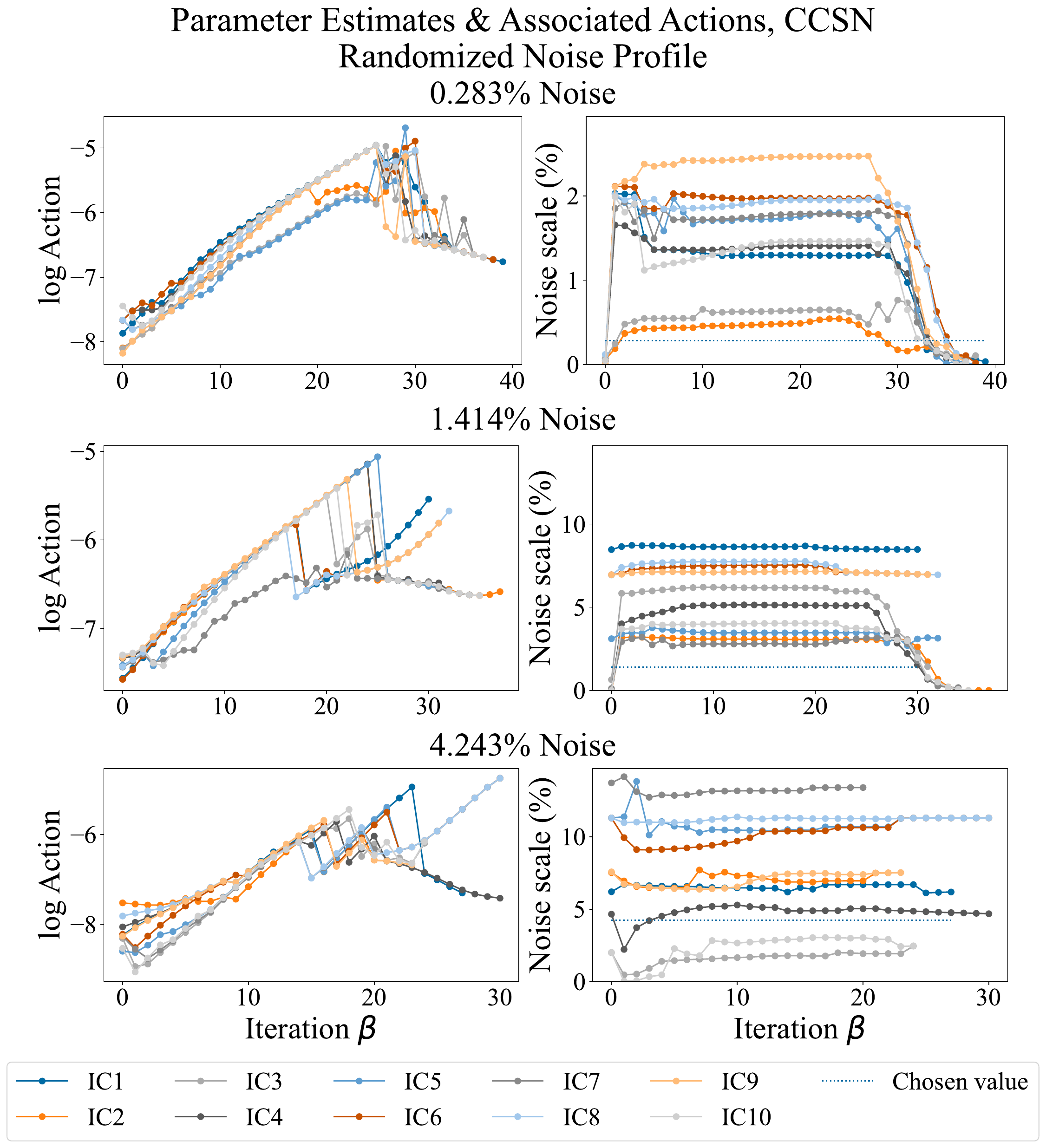}
    \caption{Results of the SDA procedure for the CCSN neutrino model, but with different randomized noise profiles provided for SDA than the ones used by forward integration to generate simulated measurements. Otherwise analogous to Fig. \ref{fig:ccsnPsurvIPOPT}.}
    \label{fig:ccsnPsurvDiff}
\end{figure} 

For CCSN neutrinos, our approach was not able to consistently infer the true noise {amplitude} used for forward integration.  However, we did see that the procedure could filter out paths which estimated substantially higher noise amplitudes than the true values, based on the evolutionary trends of the action levels at high values of $\beta$.  This held true regardless of whether or not the SDA was provided with the same noise profile shape as the forward integration.

 {In the trials with a true noise amplitude $\tilde\sigma = 0.283\%$, all paths estimated a noise amplitude close to zero, in both Figs.~\ref{fig:ccsnPsurvIPOPT} and \ref{fig:ccsnPsurvDiff}. For $\tilde\sigma = 1.414\%$,} the paths which found large estimates for $\tilde\sigma$ also showcased an exponentially increasing action at large $\beta$.  The paths which estimated noise amplitudes lower than the true values, on the other hand, experienced a dip in action {at intermediate $\beta$ values,} following which the action stopped increasing for these paths. {The action levels could thus be used to diagnose and eliminate paths with noise amplitude estimates larger than the true value.}

{For $\tilde\sigma = 4.243\%$, in the trials with the same noise profile shared between the SDA and forward procedures (Fig.~\ref{fig:ccsnPsurvIPOPT}, bottom panels), a fraction of the paths estimated the correct noise amplitude at $\beta \sim 20$, and those paths had the lowest corresponding action levels.}
{In the corresponding trials with the SDA using a different noise profile shape (Fig.~\ref{fig:ccsnPsurvDiff}, bottom panels) paths predicting either significantly higher or lower noise amplitudes than the true value each had exponentially increasing actions at large $\beta$ and could thus be filtered out.}

{The overall indication based on these trends is that our CCSN neutrino model was sensitive only to density fluctuations with sufficiently high relative amplitudes, in contrast with the solar neutrino model, which could guess correctly at lower noise amplitudes as well.}

\subsection{State Variable Estimates} \label{sec:statevar}

While the efficacy of the SDA approach was primarily assessed using the action and parameter results, we also directly compared the state variable trajectories {(i.e., evolution of the neutrino flavor polarization vector components across the radial grid)} predicted by SDA to those evaluated by forward integration.  

\begin{figure} [htb]
    \includegraphics[width=\linewidth]{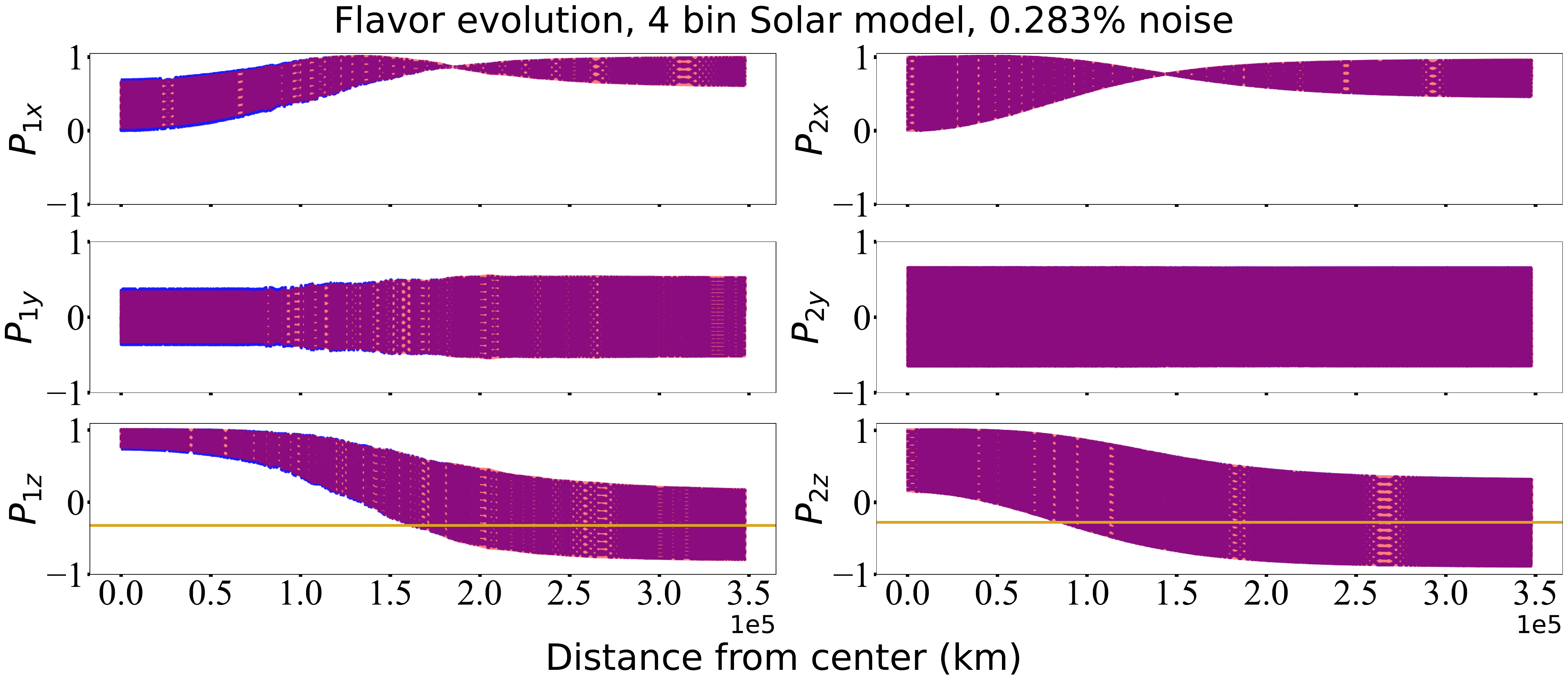}
    \includegraphics[width=\linewidth]{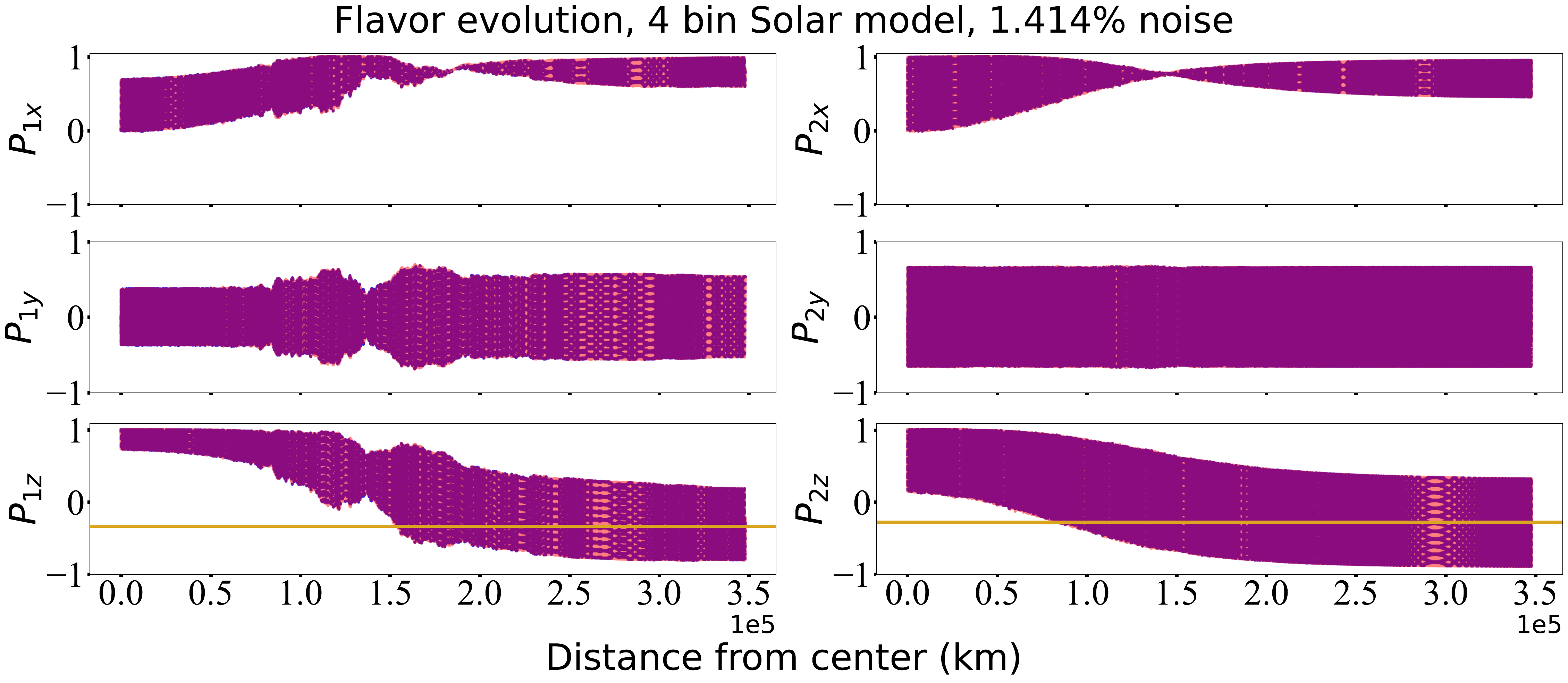}
    \includegraphics[width=\linewidth]{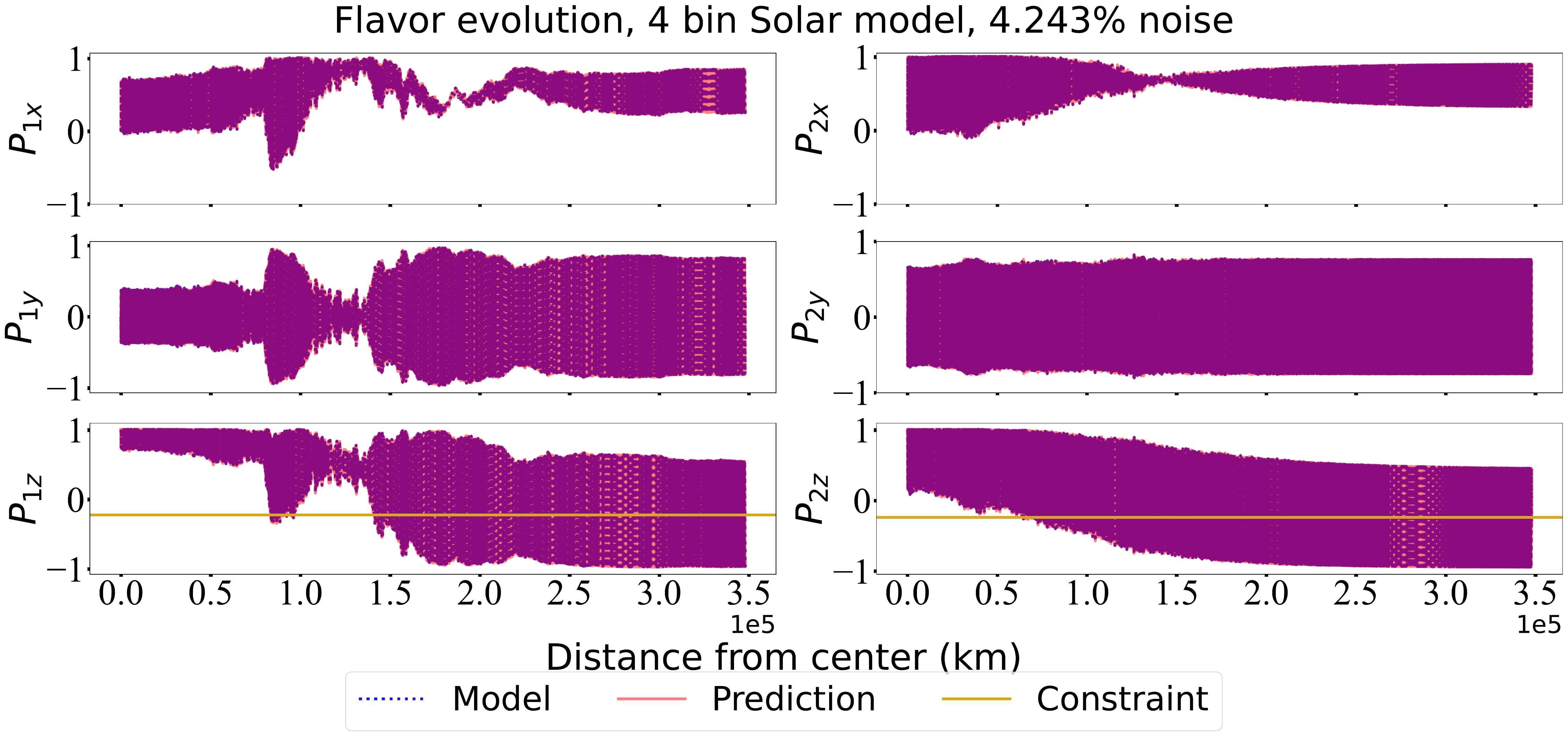}
    \caption{Flavor evolution vs distance for the solar neutrino model, in terms of the polarization vector components of the highest (left) and lowest (right) energy modes. Each panel shows the result of forward integration (``Model'') in blue, and the output of SDA (``Prediction'') in red. Also indicated in orange (``Constraint'') are average $P_z$ values in each case, corresponding to the electron neutrino survival probabilities at earth, i.e., $P_{\nu_e,\oplus}$, from Eq.~\eqref{eq:Psurv}. At each noise amplitude, we choose one path out of ten from the corresponding Fig.~\ref{fig:solar4PsurvIPOPT} panels. Within each set, we pick a path from among the ones with the lowest action levels. Specifically, the paths chosen for display here, in order of increasing noise amplitude from top to bottom, are IC1, IC1, and IC9, respectively.}
    \label{fig:solar4StateVars}
\end{figure} 
  
\begin{figure}[htb]
    \includegraphics[width=\linewidth]{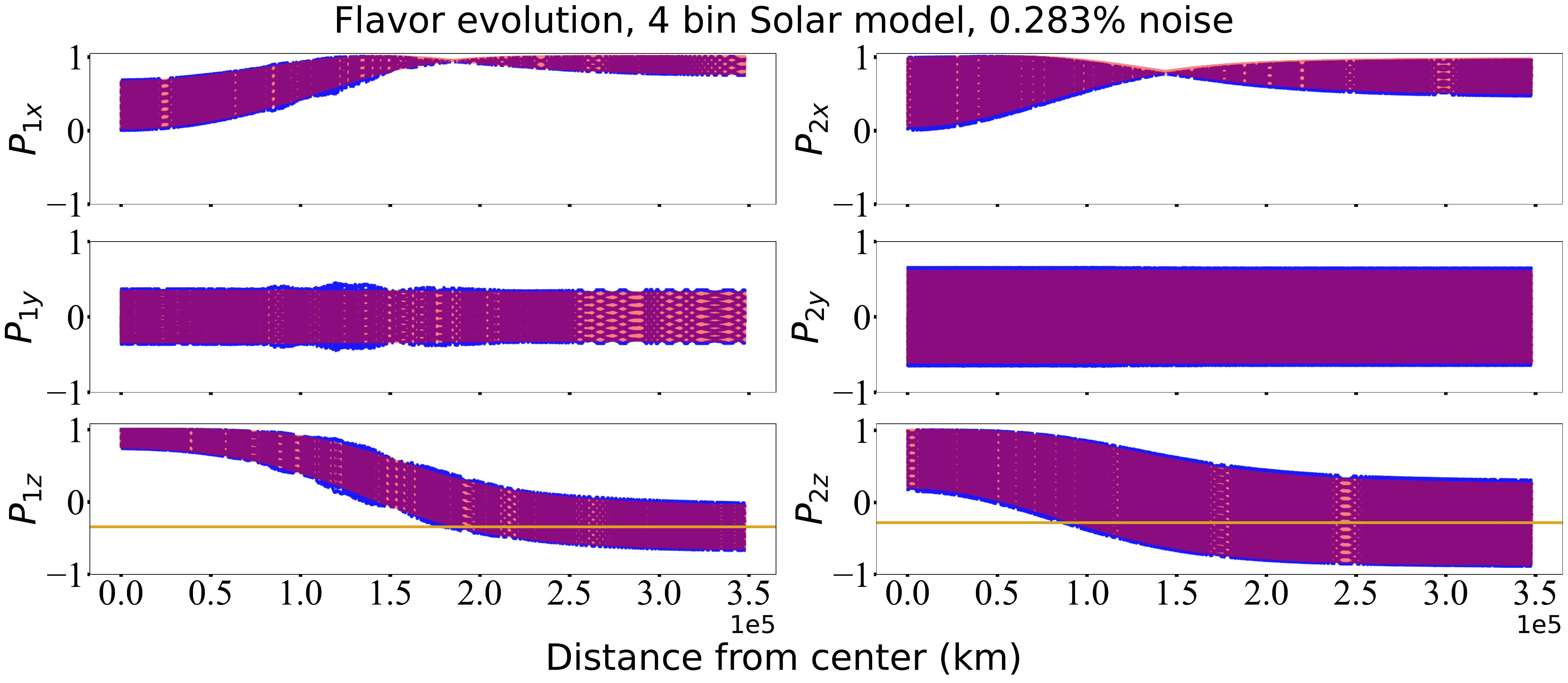}
    \includegraphics[width=\linewidth]{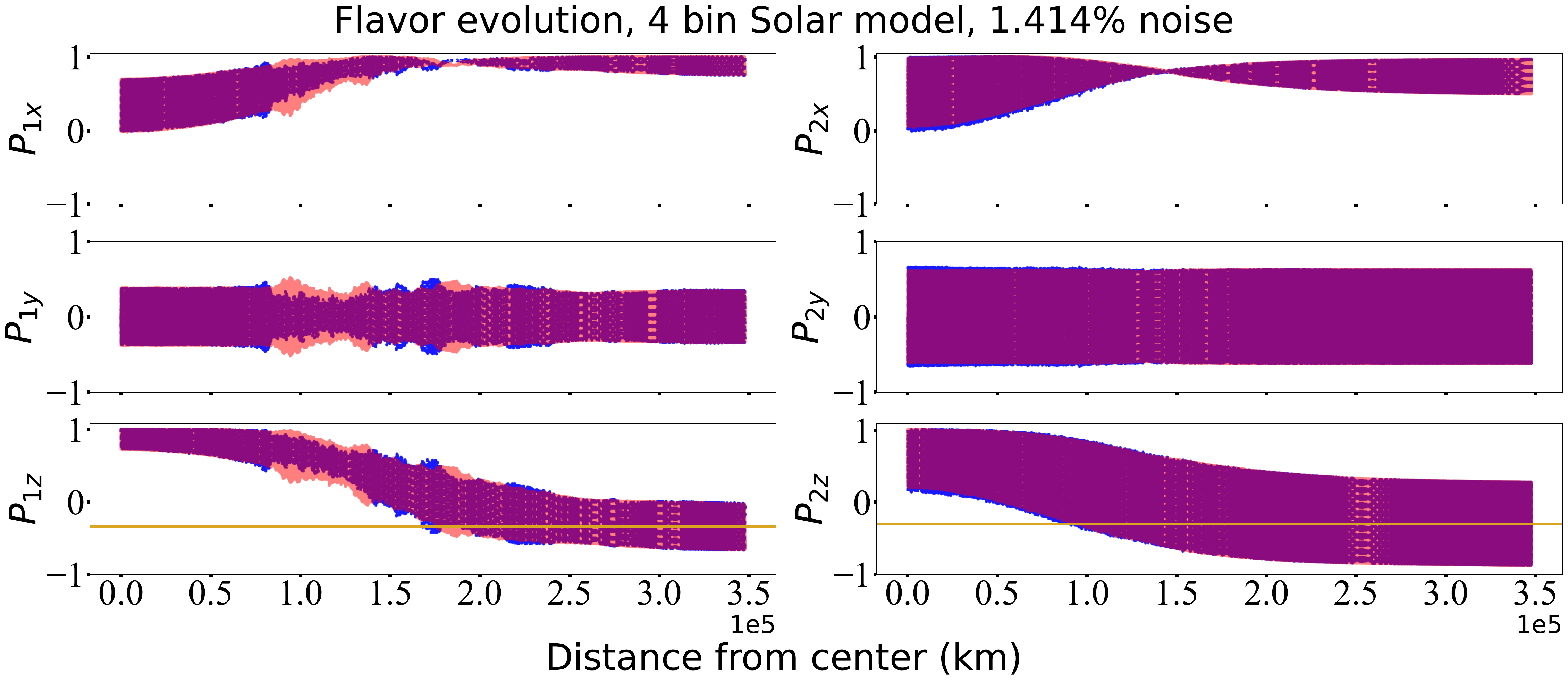}
    \includegraphics[width=\linewidth]{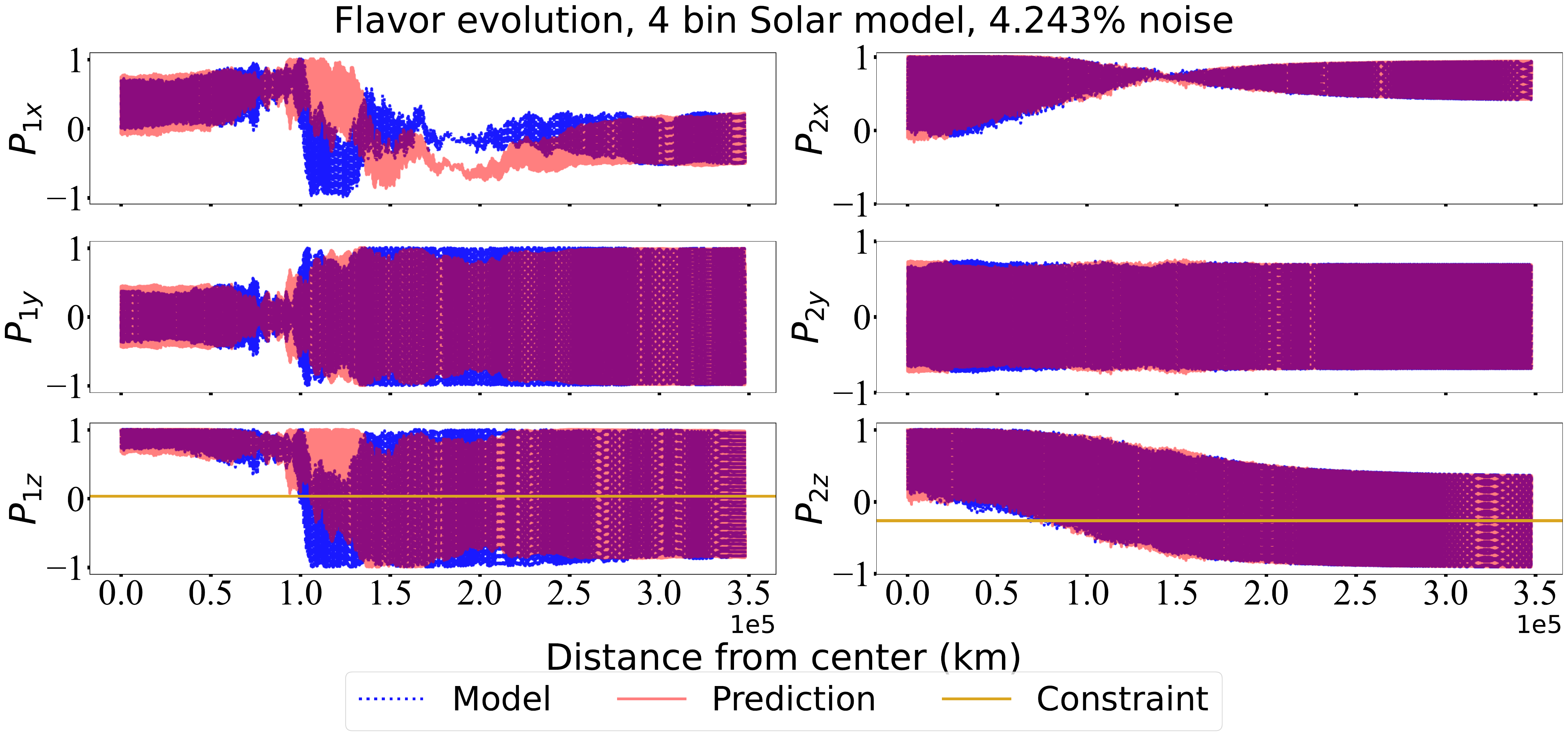}
    \caption{Flavor evolution across the radial domain for the solar neutrino model. Same arrangement as Fig.~\ref{fig:solar4StateVars}, except that in this case the SDA is given a different randomized noise profile than the forward integration. The paths chosen for display here, in order of increasing noise amplitude from top to bottom, are IC6, IC10, and IC9, respectively.}
    \label{fig:solar4StateDiff}
\end{figure}  

{Figure \ref{fig:solar4StateVars} shows one sample set of solutions from the solar neutrino calculation with the same noise profiles across the SDA and forward methods. The evolution shown here corresponds to \textit{one} path chosen from each of the left panels of Fig.~\ref{fig:solar4PsurvIPOPT}, selected from among the ones with the lowest action levels in each set. Here, we show the $P_x$, $P_y$, and $P_z$ components of the highest (left) and lowest (right) energy modes within each calculation. As with the earlier figures, the true noise amplitude values are indicated in the labels above each row of figures, and are arranged in increasing order from top to bottom. 

Figure \ref{fig:solar4StateDiff} shows the analogous results for the solar neutrino calculation with different noise profile shapes between the SDA and forward methods. In this case, unlike in Fig.~\ref{fig:solar4StateVars}, the flavor evolution pattern predicted by SDA visibly differs from the true pattern from forward integration, particularly at high noise amplitude, and most notably nearer the MSW resonance regions. This is not surprising, given that the two procedures use different noise profile shapes in this case. Crucially, the flavor evolution patterns line up much better nearer to the boundaries of the domain, where the simulated measurements are provided. This preserves the ability of SDA to predict the correct noise amplitude, even when the exact noise profile shapes differ. }

\begin{figure}[htb]
    \includegraphics[width=\linewidth]{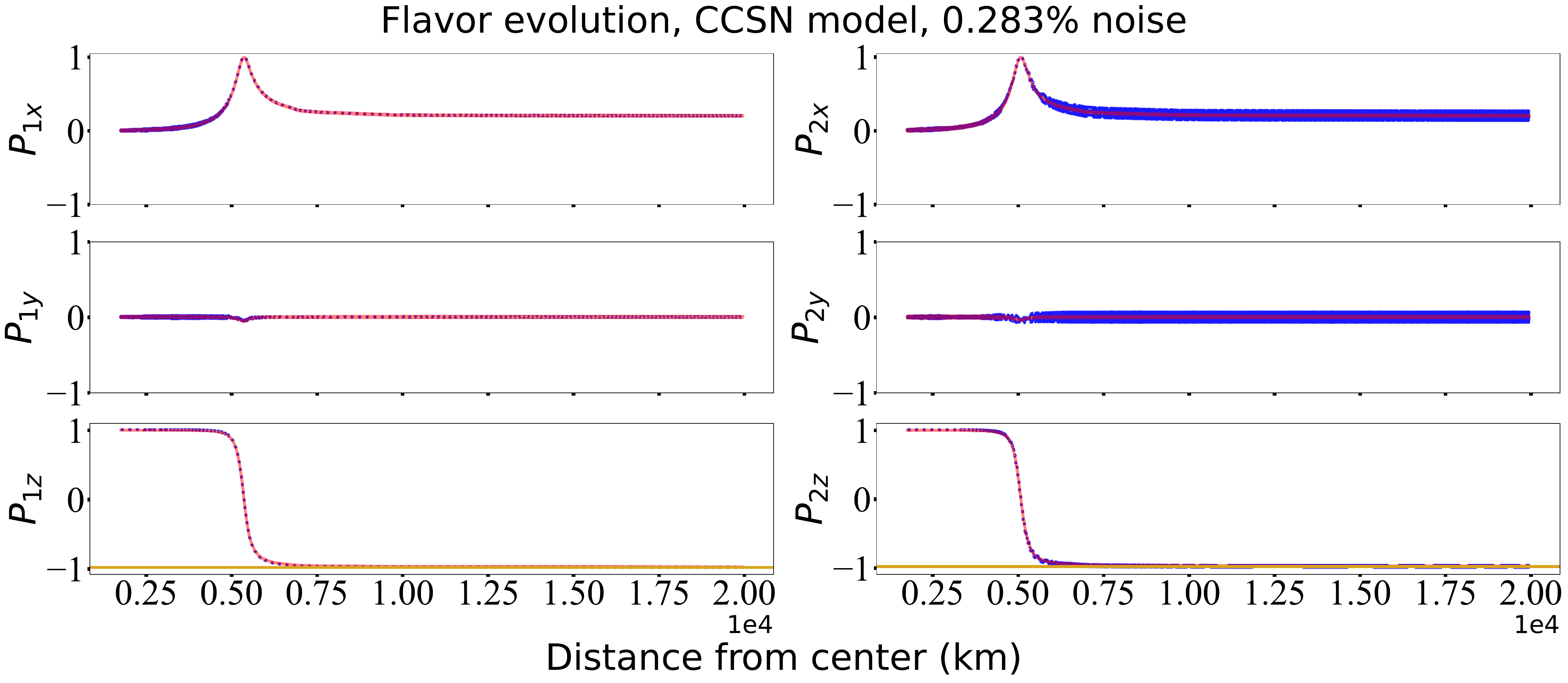}
    \includegraphics[width=\linewidth]{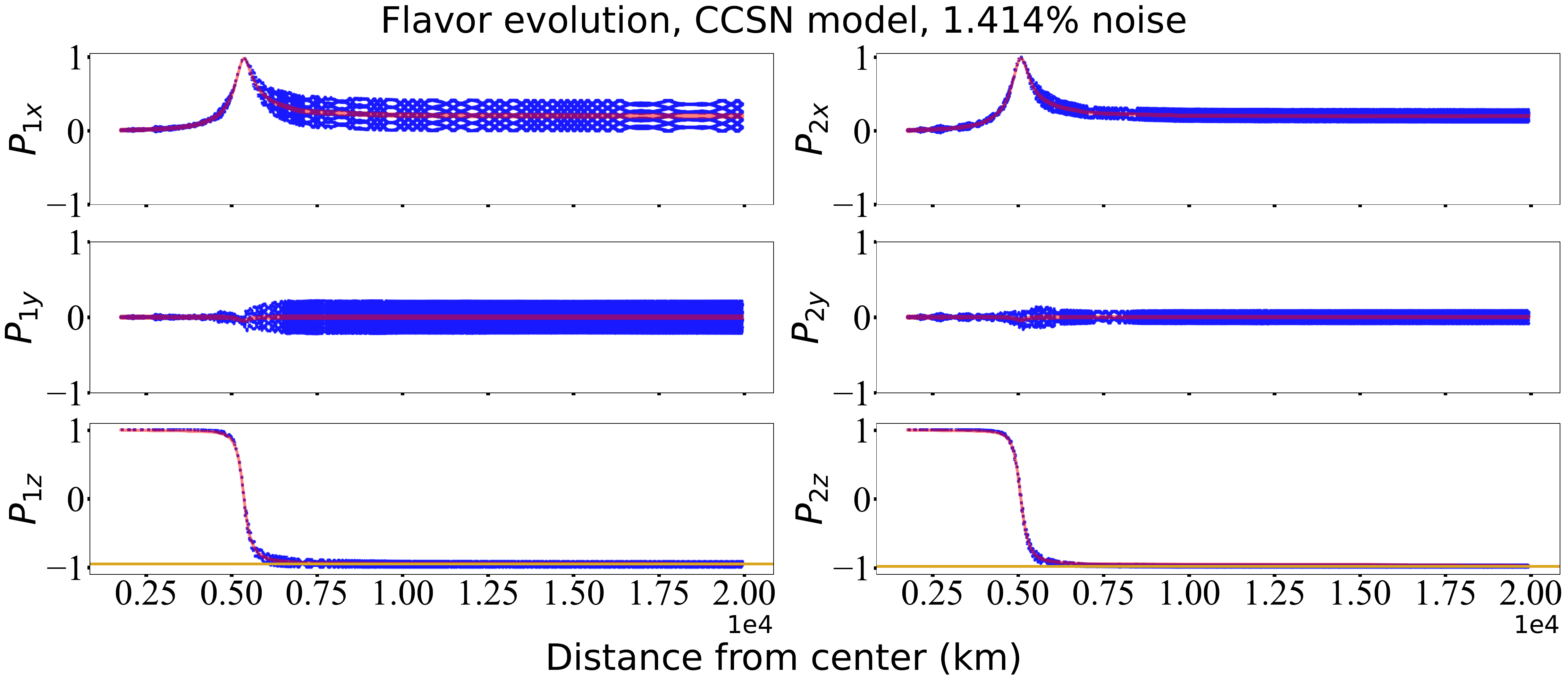}
    \includegraphics[width=\linewidth]{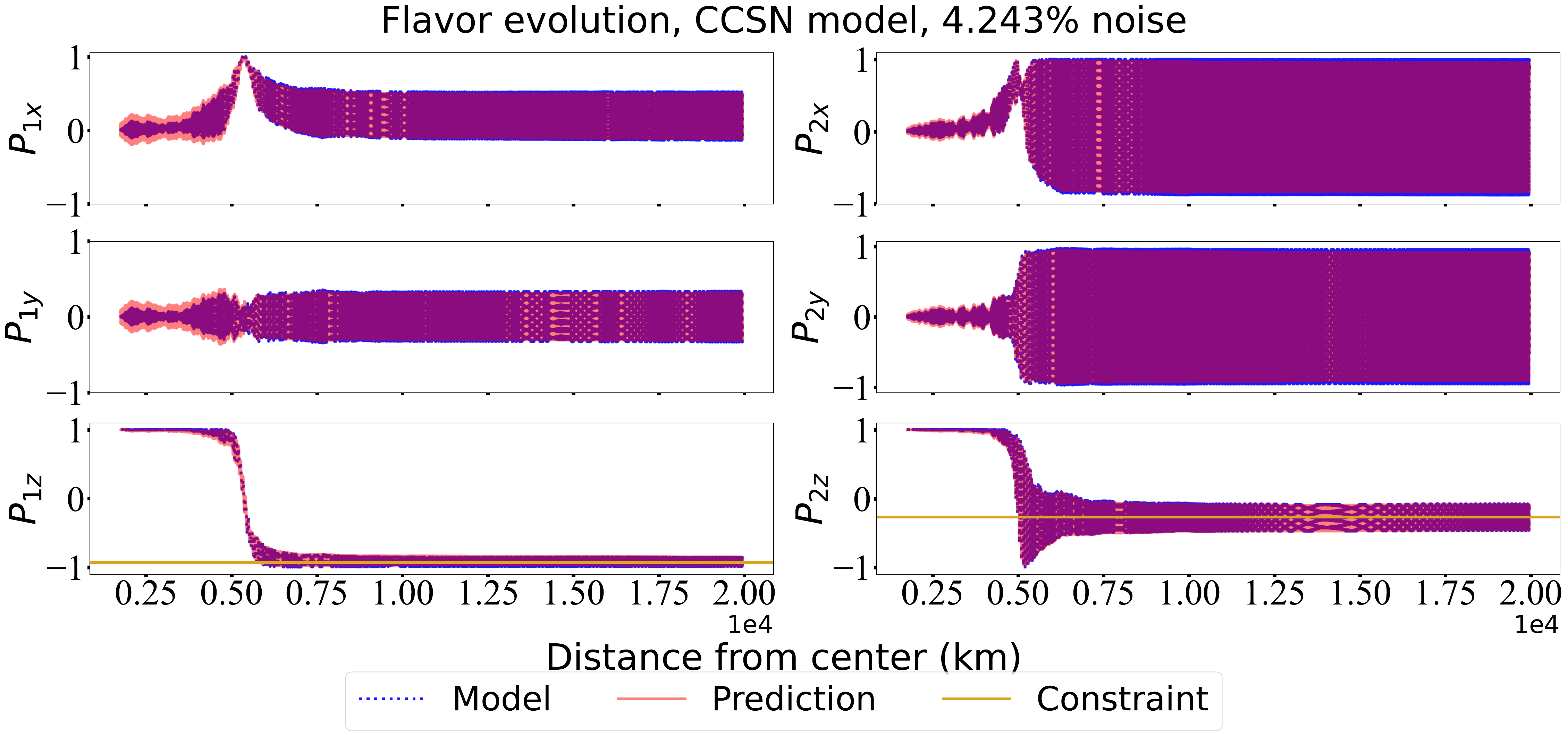}
    \caption{Flavor evolution across the radial domain for the CCSN neutrino model. Same arrangement as Fig.~\ref{fig:solar4StateVars}, and in this case, the same noise profile shape was used by the forward integration and SDA. The paths chosen for display here, in order of increasing noise amplitude from top to bottom, are IC1, IC2, and IC8, respectively.}
    \label{fig:ccsnStateVars} 
\end{figure} 

\begin{figure}[htb]
    \centering
    \includegraphics[width=\linewidth]{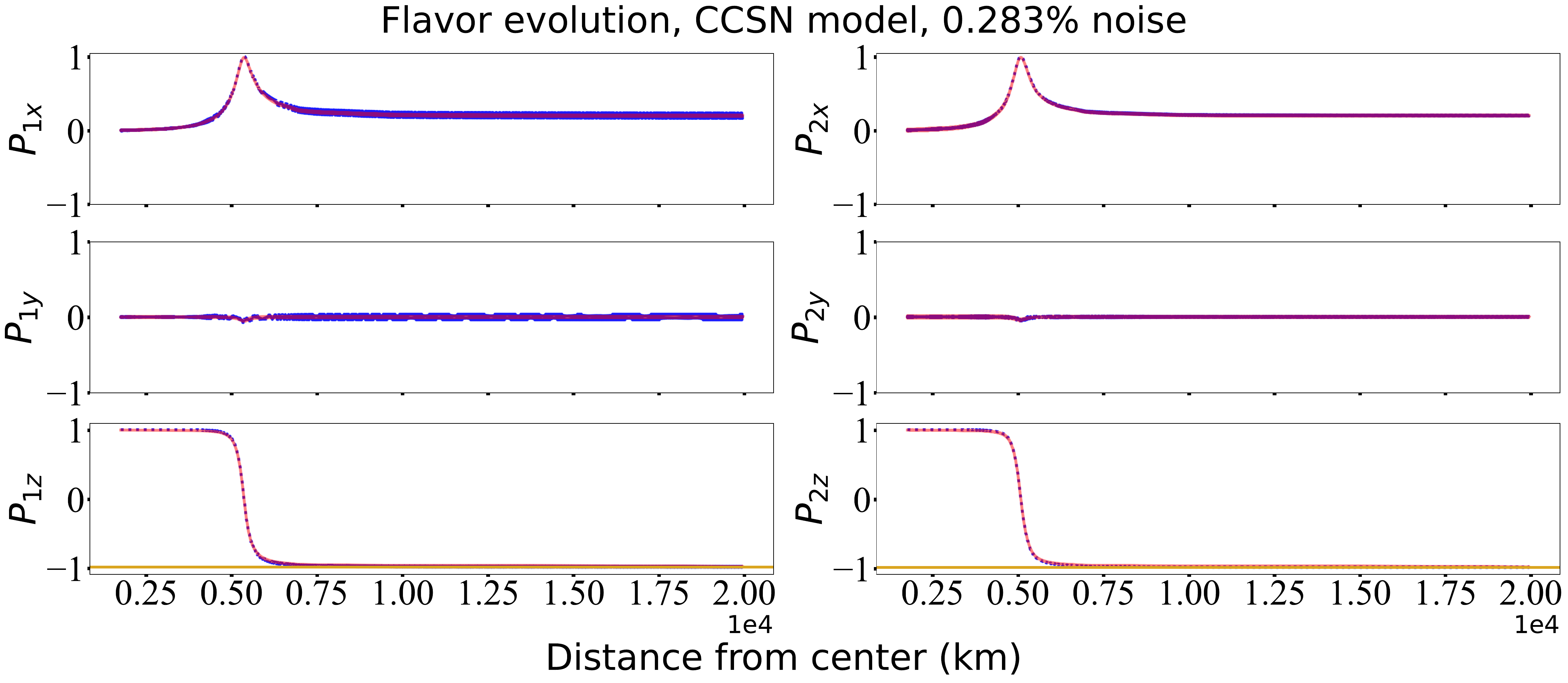}
    \includegraphics[width=\linewidth]{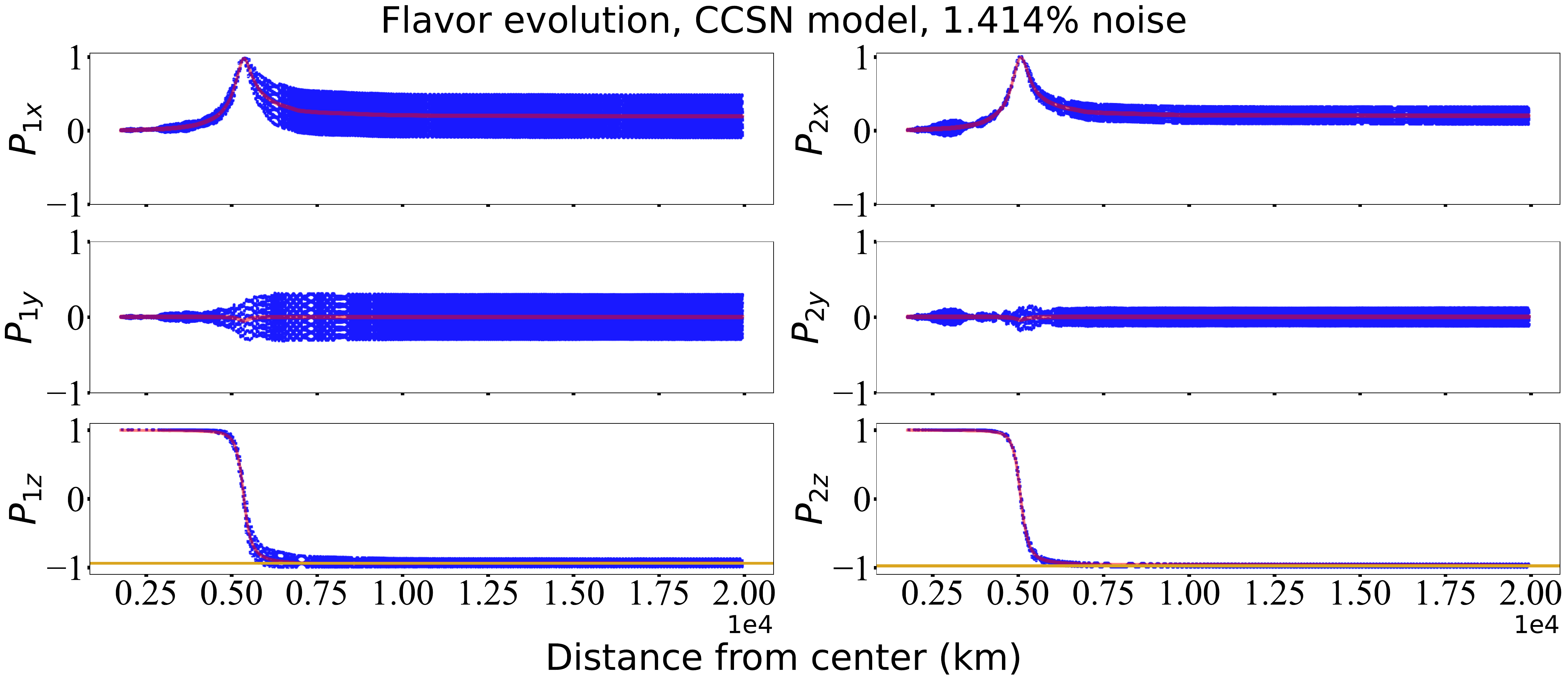}
    \includegraphics[width=\linewidth]{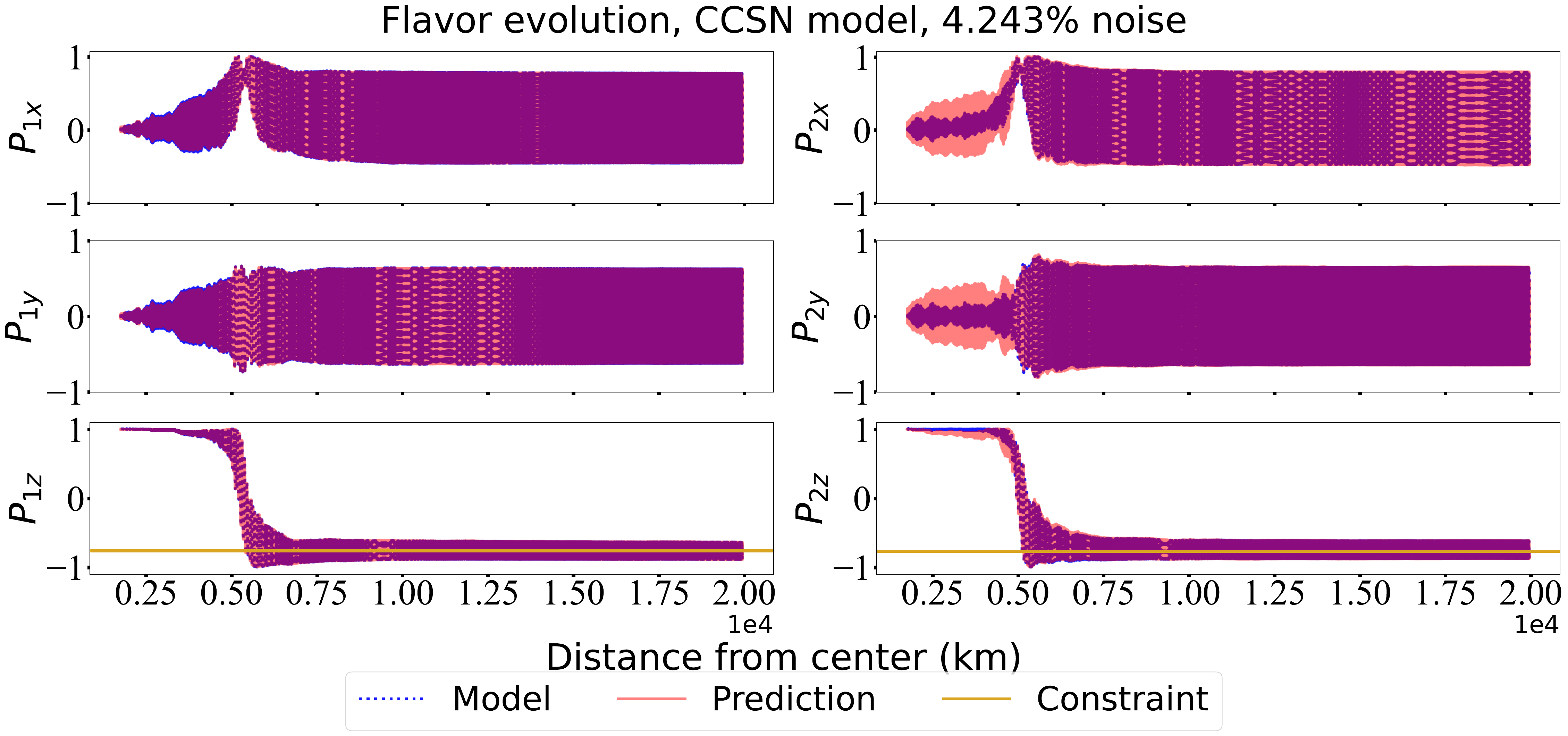}
    \caption{Flavor evolution across the radial domain for the CCSN neutrino model. Same arrangement as Fig.~\ref{fig:ccsnStateVars}, except that in this case the SDA is given a different randomized noise profile than the forward integration. The paths chosen for display here, in order of increasing noise amplitude from top to bottom, are IC2, IC2, and IC4, respectively.}
    \label{fig:ccsnDiffNoiseStateVars}  
\end{figure}

Figures \ref{fig:ccsnStateVars} and \ref{fig:ccsnDiffNoiseStateVars} show the analogous results for CCSN neutrinos, with paths chosen in each case from among the lowest action levels from each set of left panels in Figs.~\ref{fig:ccsnPsurvIPOPT} and \ref{fig:ccsnPsurvDiff}. {A close inspection of the model (forward) and predicted (SDA) flavor evolution patterns in this case reveals that the predicted flavor evolution at low and intermediate noise amplitudes exhibits a smaller oscillation amplitude than the corresponding reference models. This is consistent with the outcome that, in the trials with true $\tilde\sigma$ of 0.283\% and 1.414\%, the paths with the lowest action levels in Figs.~\ref{fig:ccsnPsurvIPOPT} and \ref{fig:ccsnPsurvDiff} each end up with a near-zero noise amplitude prediction. On the contrary, at $\tilde\sigma = 4.243\%$, the SDA flavor evolution predictions offer a closer match to those from the forward integration, again indicative of the procedure's ability to more accurately infer the correct noise amplitude when said amplitude is higher.}

\section{Conclusion} \label{sec:conclusion}

{
In summary of our results, we have demonstrated the potential to infer information about matter density fluctuations within solar and CCSN environments based on neutrino flavor observation through the use of Statistical Data Assimilation.  This approach proved more effective at high noise amplitudes and when observing neutrinos with relatively high energies, particularly those with a longer oscillation length in matter than the length scale of these fluctuations {(see Appendices \ref{app:repeat} and \ref{app:adiabaticity} for more details about this latter point).}

While the information from simulated measurements of flavor survival probabilities was not {always} sufficient to fully determine the amplitude of fluctuations modeled as white noise {(particularly in the CCSN neutrino models)}, we found that even in these cases, our SDA-based procedure could consistently exclude paths that predicted a significant higher noise amplitude than its true value. Thus, it proved capable of essentially setting an upper limit on the amplitude of these fluctuations.

This can provide insight into what observations would be needed to improve our understanding of the dynamical environments, in core-collapse supernovae in particular. {Large swathes of data from a future galactic core-collapse supernova could potentially be revelatory regarding phenomena within the CCSN environment that result in substantial irregularities in the matter density along the neutrinos' path. This proof-of-concept exploration using statistical data assimilation and a simple toy model of CCSN neutrino flavor evolution can perhaps serve as a stepping stone towards analyzing more realistic and larger-scale models using inference-based methods.}

\begin{acknowledgments} 
We dedicate this work to the memory of Eve Armstrong. 
This work was supported in part by the NSF grant No. PHY-2139004. The work of A.B.B. was also supported in part by the NSF Grant No.  PHY-2411495. 
\end{acknowledgments}

\appendix

\section{Low amplitude Noise} \label{app:lownoise}

While we primarily focused on investigating the effects of noise with a standard deviation of at least 0.283\%, we tried applying our process to smaller fluctuations as well.

To test this, we used noise with {amplitude} $\tilde\sigma = 0.071\%$, created using a Gaussian distribution with $\sigma  = 0.5\%$.  At this level, the effects on flavor evolution is small, particularly for low energy neutrinos.  This makes it unlikely that the final flavor composition of the neutrinos would be distinguishable from results with no fluctuations.

{We applied this noise {amplitude} to both solar and CCSN neutrinos, and in each case we used ten sets of random initial conditions.  The allowed search range of the noise {amplitude} parameter was $0\text{--}0.707\%$.  For each model, we saw that the SDA procedure tried out a wide range of values but was unable to distinguish between them via the resulting action, indicating that this approach is not effective for very small noise {amplitude}s.} 

\section{Repeated Forward Integration} \label{app:repeat}

As an additional metric of the effects of noise on the final state of solar and CCSN neutrinos, we performed forward integration on each setup repeatedly and recorded the final flavor compositions.  
To do this, we generated many different noise profiles with the same standard deviation, and then recorded the value of $P_{\nu_e,\oplus}$ based on $\vec{P}$ components at the last 1000 steps for each run, in accordance with Eq.~\eqref{eq:Psurv}. Then, we looked at how the distribution of these values changed as the standard deviation of the noise was increased.

Figure \ref{fig:ForwardSolar4} shows the results of this approach for our solar neutrino model.  {Not surprisingly, the variation in the $P_{\nu_e,\oplus}$ values (labeled $P_\text{surv}$ in the figure) increased with increasing noise amplitude, and the mean value across trials shifted to higher values of $P_{\nu_e,\oplus}$ with increasing noise. Similar behavior was seen in the corresponding calculations for the CCSN neutrino model, shown in Fig.~\ref{fig:ccsnPsurvF}. One surprising pattern that emerged in the solar case was that the 7 MeV neutrinos showed very little variation or upward trend in $P_{\nu_e,\oplus}$, even at the highest noise amplitudes. We further comment on this peculiar occurrence in Appendix~\ref{app:adiabaticity}.}

\begin{figure*}[htb]
    \centering
    \includegraphics[width=\textwidth]{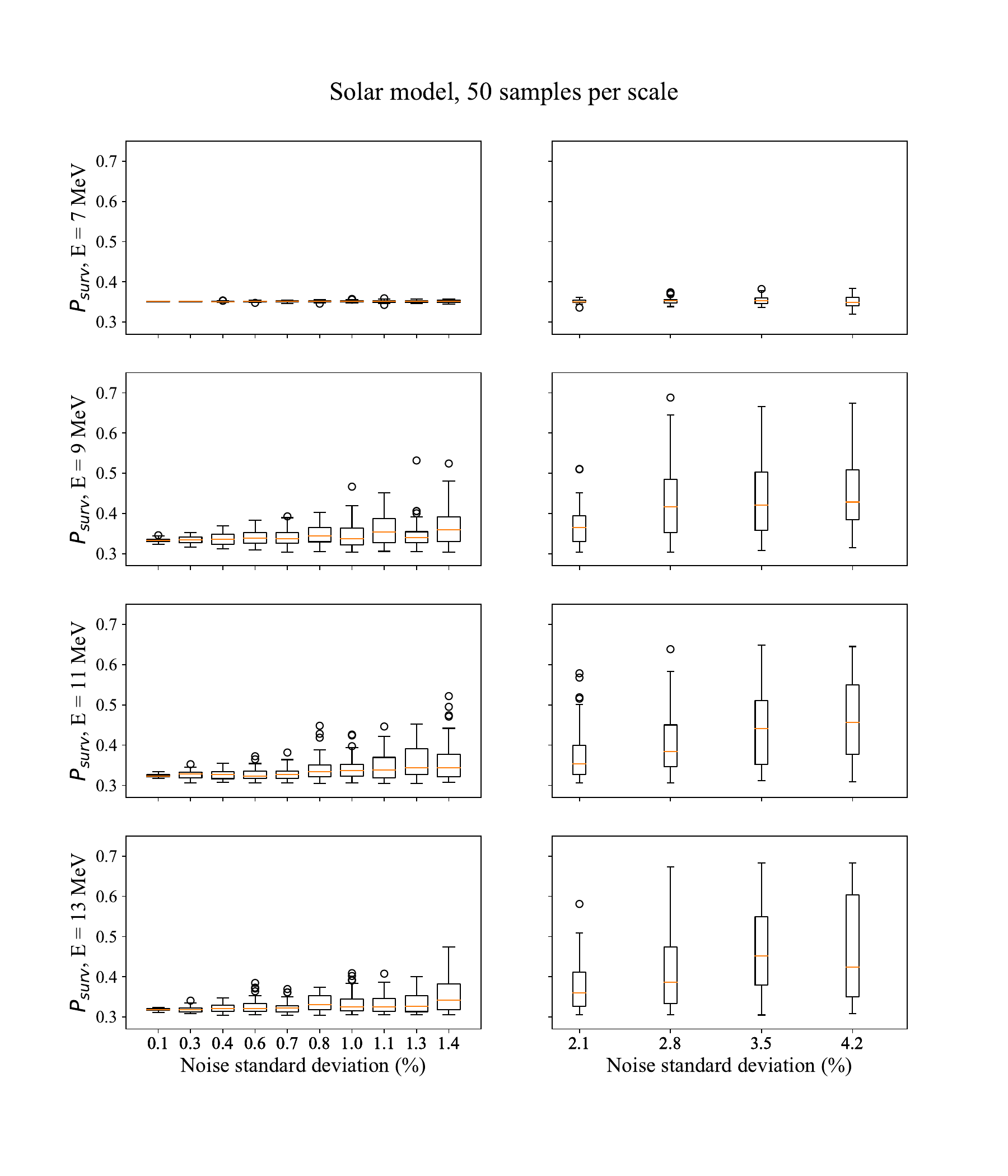}
    \caption{Electron neutrino survival probabilities for the different energy modes in the solar model, averaged over the last 1000 steps of forward integration, as a function of noise amplitude. The left and right panels in each row show relatively lower and higher noise amplitudes, respectively. Forward integration was performed 50 times for each noise amplitude, with randomly generated noise profiles each time.  Here, the orange bars represent the median value and the edges of the boxes represent the first and third quartiles.  The whiskers extend from the box to the furthest value within $150\%$ of the inter-quartile range, and dots represent any data points outside of that.} 
    \label{fig:ForwardSolar4}
\end{figure*} 

\begin{figure*}[htb]
    \centering
    \includegraphics[width=\textwidth]{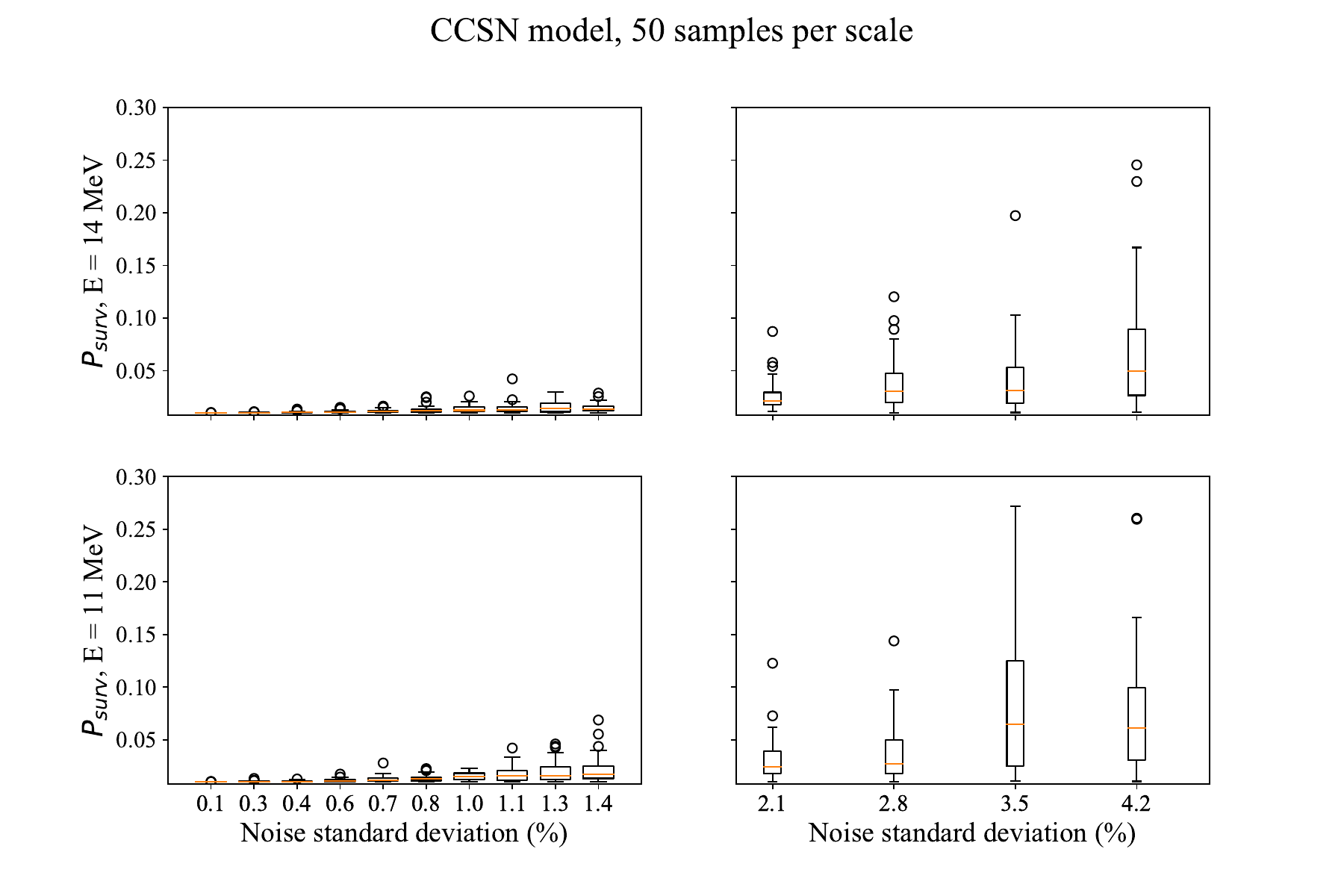}
    \caption{Same as Fig.~\ref{fig:ForwardSolar4}, but for the CCSN neutrino model.}
    \label{fig:ccsnPsurvF}
\end{figure*} 

\section{Adiabaticity and noise length scales} \label{app:adiabaticity}

An important consideration for neutrino evolution under the MSW effect is whether the matter density varies slowly enough that the overall flavor evolution happens adiabatically. The more adiabatic the flavor evolution, the less the final state would depend on the details of the matter profile, and would therefore be less sensitive to fluctuations in those cases.

As a simple heuristic for gauging the speed of this variation, we can compare the scale height $H$ of the matter density, defined as $H = [d \log n_e/dr]^{-1}$, with the in-medium oscillation length $L_\text{osc} = 4\pi E_\nu/\delta m^2_\text{eff}$ of neutrinos in matter. Here, $\delta m^2_\text{eff}$ is the effective in-medium mass-squared difference, given by
\begin{equation}
\delta m^2_{\text{eff}} = \sqrt{\left(2\sqrt{2} G_F E_\nu n_e-\delta m^2 \cos2\theta\right)^2 + \left(\delta m^2\sin2\theta\right)^2}
\end{equation}

Adiabatic flavor evolution requires $H \gg L_\text{osc}$, meaning that the matter density should change negligibly over the course of an oscillation cycle.

\begin{figure*}[htb]
    \centering 
    \includegraphics[width=\textwidth]{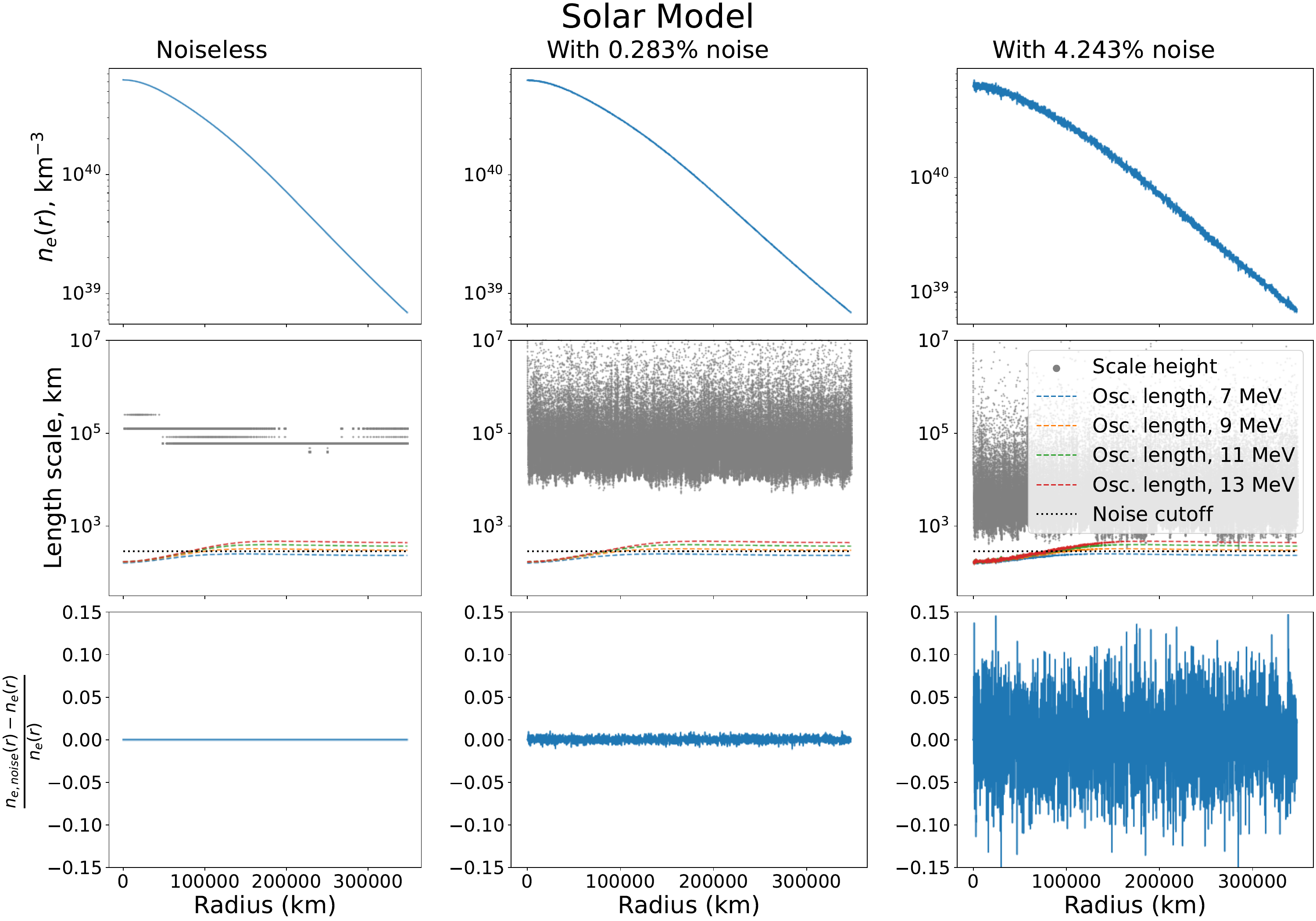}
    \caption{Illustrations of the electron density profiles (top row), the relative noise profile shapes (bottom row), and a comparison of various relevant length scales (middle row), for the solar neutrino model. The panels from left to right are in increasing order of noise amplitude.  Shown in each panel of the middle row are the scale height of the electron density (gray dots), the effective oscillation lengths of neutrinos in each mode (colored dashed lines), and the cutoff length scale of the filtered noise (black dotted line). In the solar case, all neutrinos start with faster oscillations than this noise cutoff, but only the oscillations of neutrinos with energy of 7 MeV remain faster throughout the evolution.}
    \label{fig:solarScaleHeight}
\end{figure*}

\begin{figure*}[htb]
    \centering
    \includegraphics[width=\textwidth]{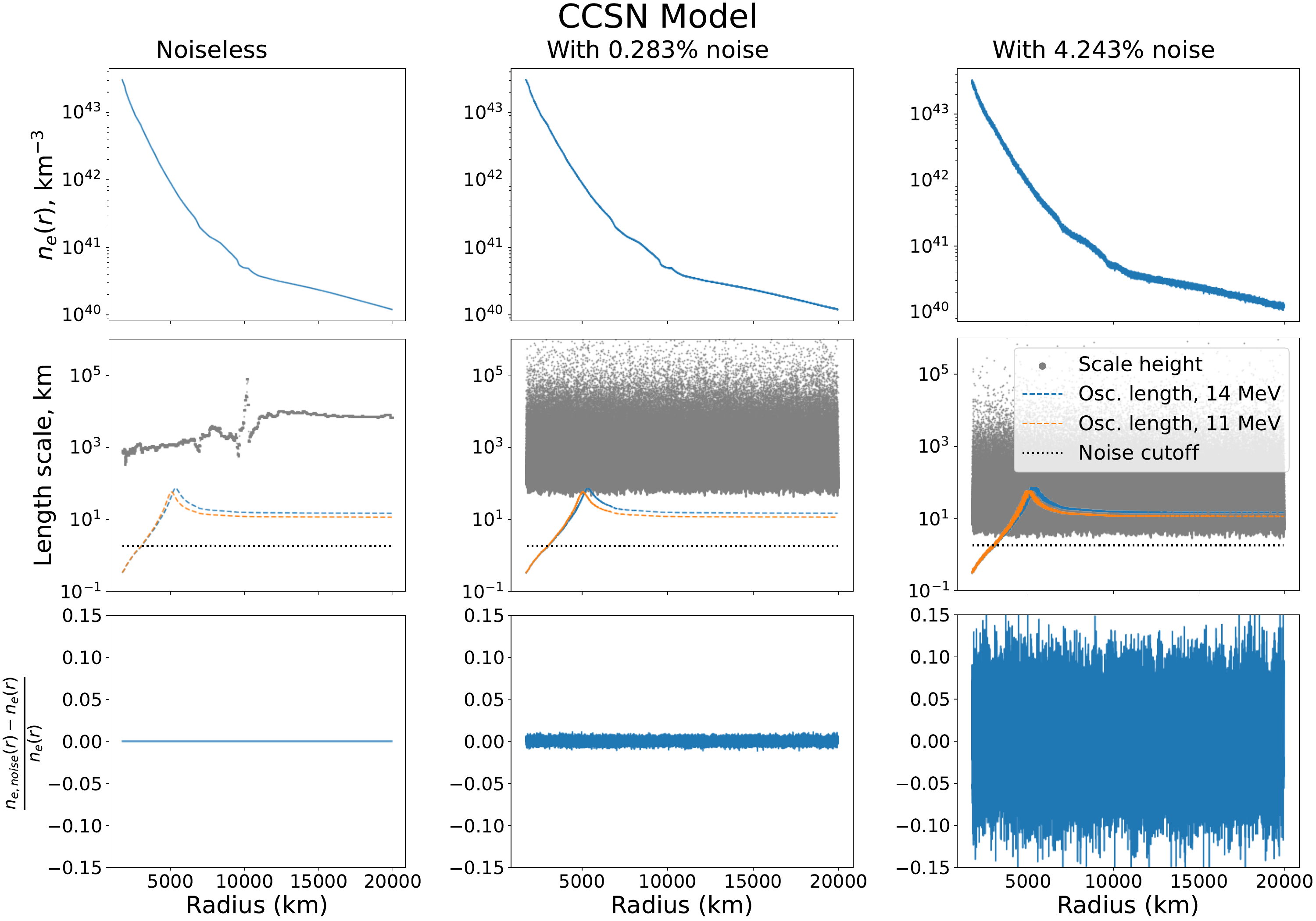}
    \caption{Same as Fig.~\ref{fig:solarScaleHeight}, but for the CCSN neutrino model.  In this case, oscillations lengths for all energy modes eventually become longer than the noise cutoff length scale.}
    \label{fig:ccsnScaleHeight}
\end{figure*}

The middle rows of panels in Figs. \ref{fig:solarScaleHeight} and \ref{fig:ccsnScaleHeight} show these quantities over the course of flavor evolution in the solar and CCSN neutrino models, respectively. {Also shown in these figures are the electron density profiles (top rows), and the relative variations in the density (bottom rows). Additionally, in the middle-row panels, a horizontal dotted line indicates the length scale associated with the noise cutoff (100 steps) in each model.} In each figure, we show the noiseless case (left column), as well as the ones with the lowest (middle column) and highest (right column) noise amplitudes considered.  

For both the solar and CCSN neutrino models, with no noise included, we see $H \gg L_\text{osc}$ for the entire region, making the evolution completely adiabatic. For the solar model, this continues to be the case at the lowest noise amplitude ($\tilde\sigma = 0.283\%$). At the highest noise amplitude ($\tilde\sigma = 4.243\%$), $H$ begins to approach $L_\text{osc}$, resulting in adiabatic evolution beginning to break down, particularly for neutrinos with higher energies.

For the CCSN neutrinos in the $\tilde\sigma = 0.283\%$ case, near the MSW region where $L_\text{osc}$ peaks, and we briefly attain $L_\text{osc} \gtrsim H$, resulting in a brief loss of adiabaticity over a narrow region near the resonance. For the highest noise {amplitude}, the domain over which adiabaticity is lost becomes a lot broader, although for small radius we still have $H > L_\text{osc}$.

{Finally, we return to the solar neutrino model to make another remark pertinent to the observation in Appendix \ref{app:repeat} of the 7 MeV energy mode lacking sensitivity to the density fluctuation amplitude. This appears to be related to the hierarchy between the oscillation length of that mode and the noise cutoff length scale. Fig. \ref{fig:solarScaleHeight} showed that our chosen noise cutoff scale ended up always larger than $L_\text{osc}$ for the 7 MeV mode throughout the radial domain explored. This was not the case for the other modes, where $L_\text{osc}$ started out lower than the noise cutoff, but eventually crossed the cutoff scale after the MSW resonance. We checked that repeating the computations of Appendix \ref{app:repeat} with a cutoff scale lower than the 7 MeV oscillation length results in that mode also acquiring greater sensitivity to the fluctuations. This is potentially an interesting outcome---indicating that neutrino flavor oscillations could be used to extract information not just about the amplitude of the noise, but also about its length scale. This warrants further exploration in a future study.}

\bibliography{bibfile}

\end{document}